\documentclass[12pt]{article}
\usepackage{makeidx}
\usepackage{multirow}
\usepackage{multicol}
\usepackage[table,dvipsnames,svgnames,x11names]{xcolor}
\usepackage{graphicx}
\graphicspath{{./Figures/}}
\usepackage[shortlabels]{enumitem} 
\usepackage{nameref} 
\usepackage{longtable, booktabs} 
\usepackage{array} 
\usepackage{makecell} 

\usepackage{pdflscape}
\usepackage{epstopdf}
\usepackage[normalem]{ulem} 
\newcommand\bluesout{\bgroup\markoverwith{\textcolor{blue}{\rule[0.5ex]{2pt}{1pt}}}\ULon}

\usepackage{amsmath}
\usepackage{bm} 
\usepackage{amssymb}
\usepackage{setspace} 
\usepackage{upgreek} 

\usepackage[utf8]{inputenc} 
\usepackage[T1]{fontenc}

\usepackage{stackengine,scalerel}
\newcommand\altPhi{\ThisStyle{\ensurestackMath{
			\stackengine{-.6\LMpt}{%
				\stackengine{-.7\LMpt}{\SavedStyle\Phi}{\rule{.5\LMex}{.7\LMpt}\kern.01ex}
				{U}{c}{F}{F}{S}}%
			{\rule{.5\LMex}{.7\LMpt}\kern.01ex}{O}{c}{F}{F}{S}}}%
}
\usepackage{float}   

\usepackage{geometry} 
\usepackage{framed}    
\usepackage{scalefnt}
\usepackage{alltt}
\usepackage[linesnumbered, ruled, vlined]{algorithm2e}

\usepackage{amsthm}
\DeclareMathOperator*{\argmax}{arg\,max}

\usepackage{mathpazo} 

\usepackage{longtable, booktabs, tabularx} 
\usepackage[font=small,labelfont=bf]{caption} 
\usepackage[flushleft]{threeparttable}  
\usepackage[hang,flushmargin]{footmisc} 
\usepackage{amsmath,amsfonts,amssymb,amsthm,epsfig,epstopdf,titling,url,array}

\usepackage{natbib} 

\usepackage[bookmarks]{hyperref} 
\hypersetup{
	unicode=false,         
	pdftoolbar=true,        
	pdfmenubar=true,        
	pdffitwindow=true,     
	pdftitle={My title},   
	pdfauthor={Author},     
	pdfsubject={Subject},  
	pdfnewwindow=true,      
	pdfkeywords={keywords}, 
	colorlinks=true,        
	linkcolor=blue,         
	citecolor=blue,       
	filecolor=magenta,      
	urlcolor=blue 
}

\usepackage{comment}

\makeatletter 
\def\blfootnote{\xdef\@thefnmark{}\@footnotetext} 
\makeatother

\defcitealias{NOAAstormdata}{NOAA NCEI, 2025}
\defcitealias{NOAAstormdataprep}{NWS, 2021}
\defcitealias{NFIPclaimshandbook}{FEMA, 2024}

\title{A zero-inflated mixed-effects spatial point process for grouped storm loss data}

\author{Lisa Gao\textsuperscript{*a} \and Sébastien Jessup\textsuperscript{ab} \and Tianxing Yan\textsuperscript{a}}
\date{June 25, 2026}

\begin{document}

\maketitle

\onehalfspacing 

\begin{abstract}
    The increasing granularity of third-party weather and exposure information can allow insurers to more effectively predict weather-related losses. However, loss outcomes are often reported in spatially grouped observations, such as at the county level, so higher resolution predictors are aggregated to align with the granularity of the outcome in standard analyses. Assuming an underlying zero-inflated mixed-effects spatial point process framework for claims arising from a common storm, we derive a model for unbalanced, multivariate zero-inflated count data that incorporates rich weather and exposure predictors observed at higher spatial granularity to predict claim patterns. The model accommodates the dependence between locations affected by a common storm in the excess zeros, as well as in the joint claim counts. Using real property exposure and loss data, we emphasize the value of incorporating granular predictors to address the localized heterogeneity of storm losses. 
\end{abstract}



\blfootnote{$^{*}$Corresponding author: \href{mailto:lisa.gao@uwaterloo.ca}{lisa.gao@uwaterloo.ca}}
\blfootnote{$^{a}$Department of Statistics and Actuarial Science, University of Waterloo, Waterloo, Ontario, N2L 3G1, Canada}
\blfootnote{$^{b}$The Co-operators, Montr\'{e}al, Qu\'{e}bec, Canada}

\newpage

\doublespacing

\section{Introduction}\label{sec:Introduction}

As the largest driver of insured catastrophe losses in Canada and the United States in recent years, severe convective storm perils have been a rising concern for property insurers \citep{aon2025climate}. 
Accurate storm loss prediction is further complicated by the highly localized and heterogeneous nature of storms. 
On one hand, technological advances in data collection provide increasingly detailed hazard and exposure information that can improve predictions. 
However, data on more sensitive claim outcomes of interest are often available on a more spatially grouped basis, for example, as the number of claims in a county or region. 
Standard analyses would involve aggregating the predictors to match the level of the response through a summary such as the average or sum in a region, at the loss of potentially valuable information, especially when storm losses exhibit substantial spatial heterogeneity. 

In this paper, we present a method for insurers to incorporate densely measured weather and exposure information to predict zero-inflated, spatially grouped losses from severe convective storms. 
We assume a zero-inflated mixed-effects spatial point process as the underlying generating process for latent claim locations arising from a storm, and derive the corresponding distribution for the observed spatially grouped claim outcomes as a multivariate zero-inflated count distribution. 
The resulting framework retains granular predictor information, accommodates excess zeros and dependence within a storm, and provides likelihood-based inference for spatially misaligned grouped loss outcomes.

Severe convective storms, which may involve thunderstorms, hail, wind, or tornados, have traditionally been deemed a ``secondary peril'' by the insurance industry due to their relatively higher frequency and lower severity compared to primary perils with the highest loss potentials and societal impact \citep{swissre2021natural}. At the individual storm level, loss potential can vary greatly, where many storms may produce little or no observed claims, while some storms generate large losses. Overall, however, the 
cumulative impact over a high frequency of heterogeneous storm events is substantial, and severe convective storms have risen to become the largest driver of catastrophe losses in recent years \citep{aon2025climate}. Insured losses averaged over \$25 billion USD in the past decade, and over \$50 billion USD in the last 3 years. Losses continue to escalate, driven primarily by increasing property exposure, urban sprawl, and rising replacement costs and inflation. In addition, while the frequency of the storms has been relatively stable, weather patterns and exposure areas prone are potentially changing with the climate \citep{gallagher2023scs}. Thus, the prediction challenge is complicated by the heterogeneity within a storm at a local, granular level, as well as the variability across different storms \citep{hornack2025developing}. 

From a data perspective, continual improvements in the availability and granularity of information on weather hazards and property exposures offer opportunities to more effectively predict storm property losses. 
Insurers can access a wealth of information from public sources, internal risk engineering or surveying teams, and third-party data and model vendors.  
For example, among rapidly-developing sources harnessed by insurers, proprietary geocoded property exposure data \citep{verisk2023geopin}, satellite and aerial imagery \citep{iii2020scs}, street-view imagery \citep{blierwong2024}, high-resolution weather radar maps \citep{shi2022assessing} and indices \citep{rogo2025italian}, and downscaled climate and weather predictions \citep{gallagher2024climate, lyubchich2019insurance}, all provide granular, densely measured input information well-suited for addressing the heterogeneity in severe convective storm losses.

However, information on insurance claims is often available more coarsely as spatially aggregated, or areal, count data, whether for policyholder confidentiality considerations regarding sensitive location information, or when consolidating industry-wide loss experience across primary insurers.
For example, the widely-used SHELDUS is a county-level loss dataset for a variety of natural catastrophe perils in the U.S. \citep{CEMHS2025SHELDUS}, and third-party vendors aggregate loss experience from primary insurers to more spatially grouped regions \citep{catIQ2024catalogue}. 
More generally, an insurer's older internal weather property claims data may not necessarily be matched to current segmentation levels and granularity, posing a challenge for linking losses to exposures and perils \citep{AAA2025finding}. 
Aggregating the predictors to the level of the response produces a conventional rectangular dataset, but obscures the localized hazard and exposure information that is valuable for storm loss prediction. Thus, the methodological challenge is in incorporating granular predictors to model spatially grouped loss outcomes, while capturing dependence between regions affected by the same storm.

Spatial point processes model the random locations of events, as opposed to the spatial outcomes at fixed, known locations \citep{cressie2015statistics}. 
Recently, \citet{Gao2022} introduced a spatial point process framework to predict the occurrence and geographical distribution of hail property insurance claims, where point-referenced claim locations are viewed as realized replicated spatial point patterns associated with storms. 
Our work is closely related to this framework, but differs in the observed data structure and role of the point process, as well as the underlying point process specification. 
Most importantly, in our setting the realized point pattern of claim locations is not observed. Instead, the response consists of replicated multivariate count outcomes on a spatially grouped level, such as postal code or county-level claim counts. 
To connect these grouped outcomes to granular weather and exposure predictors, we specify a zero-inflated mixed-effects spatial point process as a latent claim generation model. 
The mixed Poisson point process component builds on \citet{Gao2022}, with granular spatial predictors to capture observed heterogeneity and a storm-specific random effect in the claim intensity to capture unobserved heterogeneity. 
However, our full latent process further includes a storm-specific zero-inflation component, which, conditional on the observed predictors, allows dependence within a storm to arise through both the shared excess zero component and the storm-level random effect. 
The change in the observation scheme also leads to a different inferential target. Rather than directly model observed point-referenced claim locations, we derive the induced likelihood for unbalanced multivariate count outcomes across arbitrary spatial aggregations, allowing the numbers and shapes of regions to vary across storms.

More broadly, the mismatch in spatial granularity between the predictors and the response is known in the geostatistics literature as a spatial misalignment, or spatial change of support problem \citep{gelfand2010misaligned}. 
Approaches for handling the misalignment are typically either kriging-adjacent, which rely on Gaussian and linearity assumptions for spatial interpolation that are not suitable for insurance losses, or Bayesian hierarchical models, which are complex and computationally costly \citep{gotway2002combining}. 
In contrast, storm claim counts are discrete, often zero-inflated, and dependent across regions affected by the same event, motivating a count model derived from an underlying stochastic process that is computationally tractable. 

Our paper is also related to the actuarial literature on regression-based multivariate claim frequency modeling. In this literature, dependence between claim counts has primarily arisen in low dimensions for a small number of coverage types (for example, \citet{shi2014multivariate, abdallah2016sarmanov, frees2016multivariate, bolance2019multivariate, fung2019class, tzougas2021multivariate, jeong2023multivariate}). 
Geographical dependence in claim frequency across a larger number of observation units has also been studied using count regression models with spatially correlated random effects, where fixed geographical regions exhibit a Conditional Auto-Regressive correlation structure, either at the policy-level with shared regional random effects (for example, \citet{gschlossl2007spatial} and \citet{wahl2022spatial}), or for spatially grouped regional claim counts \citep{tufvesson2019spatial}.  
In contrast to modeling individual policyholder claim frequency for ratemaking, we focus on the joint claim outcomes at the weather event level, where the dependence arises from a common storm affecting varying numbers and shapes of regions.

Our approach to addressing the spatial misalignment challenge is motivated by the underlying stochastic process. An analogous setting is in loss reserving, where triangular loss data are available on an aggregate level, and the methods used to model such data are often supported by underlying stochastic models on granular claims developments that mirror the data generating process for the aggregate data \citep{wuthrich2008stochastic, taylor2012loss}. 
In the count modeling context, our model nests the Type II multivariate zero-inflated negative binomial regression in \citet{zhang2025comparative} as a special case when the predictor and outcomes are spatially grouped at the same level. However, our unique replicated spatial data setting presents higher-dimensional and unbalanced outcomes when different storms affect different geographical regions. Since the method is derived from the underlying spatially continuous point process, it is agnostic to the definition of the regions, and cohesively handles changing spatial segmentation. 

Our work further supplements the actuarial literature on modeling severe convective storm losses. Existing studies have considered modeling the aggregate storm losses directly  \citep{hua2017factor} or modeling temporally and spatially aggregated storm loss outcomes \citep{haug2011future, huang2024storm, fung2025investigating}. 
We instead take a granular approach within a storm event to focus on specific characteristics of storm losses, including the spatial heterogeneity, zero-inflated outcomes, and dependence within an event. 
From a claims management perspective, the resulting granular risk maps can inform insurers' catastrophe-specific response actions and help coordinate efficient claims investigation activities, especially when claims handling expenses can accumulate quickly through higher frequency storm events.

Against the aforementioned literature and observed data setting, our paper makes several contributions. 
First, we address the spatial misalignment between rich weather and exposure predictors and spatially grouped discrete loss outcomes by deriving the observed multivariate count distribution from an underlying zero-inflated mixed-effects spatial point process of latent claim occurrences. 
Our method captures the localized heterogeneity in storm losses while cohesively accommodating areal count outcomes from arbitrary regions and extents of spatial aggregation. 
Second, the model captures within-storm dependence through two channels. The excess zero component depends on storm-level characteristics and allows dependence in the joint absence of claims across all regions affected by the storm, while the storm-specific random effect captures unobserved storm heterogeneity and induces dependence in the regional claim counts generated by the point process component. 
Third, we derive a computationally tractable closed-form likelihood for the observed spatially grouped loss outcomes, and develop a tailored EM algorithm for likelihood-based estimation of the multivariate zero-inflated count model. 

We study the performance and practical value of the proposed method through simulation experiments and an application to real storm loss data. 
The application uses a unique replicated spatial dataset constructed from the National Oceanic and Atmospheric Administration (NOAA) Storm Events Database, linked to granular weather and property exposure information, and illustrates the value of incorporating densely measured predictors over their aggregated counterparts.

The remainder of this paper is organized as follows: Section \ref{sec:ZISpatialPointProcess} introduces the zero-inflated spatial point process for replicated grouped data; Section \ref{sec:Estimation} describes the tailored model estimation procedure; Section \ref{sec:NumericalExperiment} conducts numerical studies to illustrate the estimation at various levels of spatial aggregation; Section \ref{sec:Data} provides an overview of the real storm, exposure and claims data; Section \ref{sec:Application} demonstrates the proposed method using the replicated storm loss data; and Section \ref{sec:Conclusion} concludes.

\section{Model} \label{sec:ZISpatialPointProcess}

We adopt a spatial point process perspective as the underlying generating process for the occurrence and geographical distribution of claims associated with a storm, from which we derive the joint distribution for the observed spatially grouped claim count data. We first introduce the underlying zero-inflated spatial point process that incorporates granular weather and exposure characteristics, followed by the corresponding model for spatially grouped outcome observations. 

\subsection{Zero-inflated mixed Poisson spatial point process}  

Consider a collection of $m$ storms, where $\bm{Y}_i$ denotes the spatial point process of claims associated with the $i$-th storm. We use $\bm{Y}_i$ to introduce our spatial point process notation, and then specify its distribution under our zero-inflated mixed Poisson model. 
$\bm{Y}_i$ is a locally finite, random countable subset of a two-dimensional space $S \subseteq \mathbb{R}^2$, i.e., it is a set of a random number of points at random locations.  
The random counting measure $N_i(B)$ counts the number of points from $\bm{Y}_i$ that fall within a bounded region $B \subseteq S$, and the distribution of $\bm{Y}_i$ is determined by the joint distribution of $N_i(B_1), \ldots, N_i(B_k)$ for any bounded Borel sets $B_1, \ldots, B_k$ in $S$ and $k \in \mathbb{N}_0$ \citep{moller2003statistical}. In practice, points are observed within a bounded spatial observation window in $S$, denoted by $W_i$ for the $i$-th storm. For example, $W_i$ may represent the affected region for storm $i$. 

In our study, we specify the spatial point process $\bm{Y}_i$ as a zero-inflated point process with mixed-effects. The 
random effect $U_i$ captures the unobserved storm-specific heterogeneity; conditional on $U_i$, the spatial point process $\bm{Y}_i$ is a zero-inflated Poisson spatial point process:
\begin{align} \label{eq:condPtProcess}
	\bm{Y}_i|U_i = \begin{cases}
		\emptyset
		, & \text{with probability } p_i, \\
		\bm{Y}_i^\ast|U_i, & \text{with probability } 1-p_i,
	\end{cases}
\end{align}
where $p_i$ is the probability of excess zero claim occurrences for storm $i$ through the empty set of points, and $\bm{Y}_i^\ast | U_i$ is a Poisson spatial point process with a multiplicative factor $U_i$ in its conditional intensity function specified below. 
Our construction is in the spirit of the standard mixture approach for zero-inflated count distributions, where here, the point process of random claim occurrences and locations within a storm has an amplified probability of no occurrences at all. 
The excess zero probabilities $p_i$ may further differ across storms and depend on various storm-level risk characteristics, for example, as a generalized linear model with a logit link so that 
\[
p_i=(1+\exp\{ -\bm{z}_i^{\top}\bm{\eta} \})^{-1},
\]
where $\bm{z}_i=(z_{i1}, \ldots, z_{id_1})^{\top}$ denotes the vector of storm-level predictors associated with the zero-inflation component, and $\bm{\eta} = (\eta_1, \ldots, \eta_{d_1})^{\top}$ is the corresponding coefficient vector.

The conditional Poisson spatial point process $\bm{Y}_i^\ast|U_i$ is characterized by its first-order intensity function $\lambda_i(s|U_i)$, a spatially varying function of location $s \in W_i$ that describes the average number of events per unit area, for an infinitesimal area around $s$. The related intensity measure $\mu_i(B|U_i) = \int_B \lambda_i(s|U_i) ds$ represents the expected number of points in a bounded region $B \subseteq S$. 
We specify an inhomogeneous Poisson spatial point process that accounts for both observed and unobserved heterogeneity through the conditional intensity
\begin{align} \label{eq:condIntensity}
	\lambda_i(s|U_i) = U_i \, \gamma \exp\{\bm{x}_i(s)^{\top}\bm{\beta}\}, \quad s \in W_i, 
\end{align}
where $\gamma>0$ is the common baseline intensity across all of the storms, and the intensity coefficients $\bm{\beta}=(\beta_1, \ldots, \beta_{d_2})^{\top}$ quantify the effect of log-linearly incorporated predictors $\bm{x}_i(s)=(x_{i1}(s), \ldots, x_{i{d_2}}(s) )^{\top}$, which may vary with location $s$, but do not need to. More generally, the baseline intensity could also be allowed to vary spatially, for example, by specifying $\gamma(s)>0$ as a spatial trend in the point process intensity \citep{baddeley2016spatial}. 
In this paper, we use the constant baseline specification $\gamma(s)\equiv \gamma$ for parsimony, and focus on the spatial heterogeneity captured through the observed predictors $\bm{x}_i(s)$. 
Conditional on $U_i$,  $\bm{Y}_i|U_i$ is an inhomogeneous zero-inflated Poisson point process so that the number of claims $N_i(B|U_i)$ in any bounded region $B$ follows a zero-inflated Poisson distribution with excess zero probability $p_i$ and Poisson mean $\mu_i(B|U_i)$. 

We assume that the storm-specific random effects $U_i,\, i=1, \ldots, m,$ are i.i.d.\ with mean 1 for identifiability, and follow a gamma distribution with scale parameter $\psi>0$ so that $\mathrm{Var}(U_i)=\psi$ captures additional variability in the intensity across storms. 
The gamma distributed random effect has been shown to flexibly characterize unobserved storm-specific heterogeneity \citep{Gao2022}, and results in tractable, closed-form expressions for many quantities of interest in the unconditional, zero-inflated mixed Poisson point process $\bm{Y}_i$. 
For example, the number of claims $N_i(B)$ in a bounded region $B$ follows a zero-inflated negative binomial distribution with excess zero probability $p_i$ and negative binomial mean $\mu_i(B) = \mathrm{E}\left[\mu_i(B|U_i)\right]$ and dispersion $\psi^{-1}$. The probability mass function is given, for $n=0, 1, \ldots,$  by
\begin{align*}
	&\Pr(N_i(B) = n) = \left[ p_i + (1-p_i) \left( \frac{\psi^{-1}}{\psi^{-1} + \mu_i(B)} \right)^{\psi^{-1}} \right]^{I(n=0)} \times \\
	&\qquad \qquad \qquad \qquad \qquad \left[ (1-p_i) \frac{\Gamma\left( \psi^{-1} + n  \right) }{\Gamma\left(\psi^{-1}\right) n! } \left( \frac{\psi^{-1}}{ \psi^{-1} + \mu_i(B) } \right)^{\psi^{-1}} \left( \frac{\mu_i(B)}{\psi^{-1} + \mu_i(B)} \right)^n \right]^{I(n>0)}.
\end{align*}
As the underlying stochastic generating process of claim occurrences, the model $\bm{Y}_i$ describes the random occurrence locations of claims for storm $i$, and accommodates several notable features of storm-related losses. 
First, the effect of highly localized observed heterogeneity within and across different storms is incorporated through the rich, granular spatially varying weather information and property characteristics associated with a storm $\bm{x}_i(s)$. 
Second, the storm-specific random effects $U_i$ further capture the dependence between geographical regions affected by the same storm that result from unobserved heterogeneity. 
Third, joint zero claims at the storm level depend on various observed storm characteristics $\bm{z}_i$, as well as on the granular predictors $\bm{x}_i(s)$ and unobserved storm-specific heterogeneity from $U_i$. 
Next, retaining these model characteristics in the underlying claim occurrence, we flexibly and cohesively accommodate the more aggregated loss outcomes. 

\subsection{Grouped outcomes under the spatial point process}  

Considering the spatially grouped nature of observed claims data, we are interested in obtaining the joint distribution of claim counts in different regions from the underlying zero-inflated mixed Poisson spatial point process of claims.
In particular, for any disjoint, bounded regions $B_1, \ldots, B_k \subseteq W_i$, the number of claims $(N_i(B_1), \ldots, N_i(B_k))$ from storm $i$ jointly follow a multivariate zero-inflated negative binomial distribution with negative binomial component mean parameters $(\mu_i(B_1), \ldots, \mu_i(B_k))$ and dispersion parameter $\psi^{-1}$:
\begin{align}
	& \Pr(N_i(B_{1})=n_{i1}, \ldots, N_i(B_{k})=n_{ik})  \nonumber \\
    & \qquad = \left[  p_i + (1-p_i) \left(\dfrac{\psi^{-1}}{\psi^{-1} + \mu_{i+}} \right)^{\psi^{-1}}  \right]^{I(  n_{i+} = 0 )} \times \nonumber \\
    & \qquad \qquad \left[ (1-p_i) \dfrac{\Gamma(\psi^{-1} + n_{i+})}{\Gamma(\psi^{-1}) \prod_{j=1}^{k} n_{ij}!}\ \left(\dfrac{\psi^{-1}}{\psi^{-1} + \mu_{i+}} \right)^{\psi^{-1}} \prod_{j=1}^{k} \left( \dfrac{\mu_i(B_j)}{ \psi^{-1} + \mu_{i+} } \right)^{n_{ij}} \right]^{I( n_{i+} > 0 )}, \label{eq:ZInegmult}
\end{align}
where $n_{i+} = \sum_{j=1}^{k} n_{ij}$, $\mu_{i+} = \sum_{j=1}^{k} \mu_i(B_j)$, and $p_i \in (0, 1)$ is the inflated probability of joint zero claims across all of the regions. The above result follows from Property 2 of the mixed-effects Poisson point process in \cite{Gao2022}, where the specific form of the multivariate negative binomial component in \eqref{eq:ZInegmult} is the negative multinomial distribution, the classical multivariate generalization of the negative binomial distribution. 
We use the term multivariate negative binomial to align with the actuarial multivariate count regression literature, where the term more broadly encompasses multivariate count models with negative binomial margins \citep{shi2014multivariate}.

The model (\ref{eq:ZInegmult}) provides a joint distribution for the observed spatially grouped claim counts $(n_{i1}, \ldots, n_{ik})$ from regions $B_1, \ldots, B_k$ in storm $i$ that incorporates the more densely measured weather and exposure information within the underlying claims generating process. Thus, we retain the inclusion of rich predictors in the joint zero counts and claim frequency, unobserved storm-specific heterogeneity, and dependence within a common storm. 

We further emphasize a few points about this multivariate zero-inflated count model. 
First, the negative binomial component mean parameters $\mu_i(B_j), j=1\ldots, k,$ come from the marginal intensity measure of the point process component. 
Since $\mathrm{E} \left[ U_i \right]=1$, we obtain $\mu_i(B_j) = \mathrm{E}\left[ \mu_i(B_j | U_i) \right]  = \int_{B_j} \gamma \exp\{ \bm{x}_i(s)^{\top} \bm{\beta} \} ds$ as a spatial integral  over the region $B_j$ involving the spatially varying predictors. 
The model therefore nests the Type II multivariate zero-inflated negative binomial regression model \citep{zhang2025comparative}, which has dependence among the margins through the zero-inflation and the shared gamma mixing variable, as a special case when the predictors $\bm{x}_i(s)$ are also spatially grouped at the same level as the claim outcomes.
Second, because the joint model for $(N_i(B_1), \ldots, N_i(B_k))$ is derived from the underlying zero-inflated mixed Poisson spatial point process $\bm{Y}_i$, the method is agnostic to the definition of the disjoint geographical regions $B_1, \ldots, B_k$. Thus, the model cohesively addresses different levels of granularity of the observed data (e.g., at the county or postal code level) from different segmentation, as well as changing dimensions of the multivariate count distribution from different storms. 
Third, the gamma distributed random effect balances flexibility and tractability in computation and simulation; however, our underlying method of deriving a multivariate count distribution from a zero-inflated doubly stochastic spatial point process to handle grouped loss data can be generalized, albeit at much higher computational cost.

\section{Model estimation}\label{sec:Estimation} 

Unlike typical mixed-effects regression models, the zero-inflated mixed Poisson spatial point process model derived for grouped data has a tractable, closed-form log-likelihood function, leading to more computationally convenient likelihood-based parameter estimation. 
Assume that we observe $n_{ij} \geq 0$ claims from $B_{ij}$, the $j$-th subregion of storm $i$, for $j=1,\ldots,k_i$ and $i=1,\ldots,m$. 
For storm $i$, let $\bm{n}_i = (n_{i1}, \ldots, n_{ik_i})^{\top}$ be the observed joint claim counts from subregions $B_{i1}, \dots, B_{ik_i}$, $\bm{z}_i=(z_{i1}, \ldots, z_{id_1})^{\top}$ 
be the observed storm-level covariates related to the joint zero-inflation probabilities, and $\bm{x}_i(s) = (x_{i1}(s), \ldots, x_{id_2}(s))^{\top}$ be the weather and exposure characteristics that may vary spatially, but may also include non-spatially-varying predictors from $\bm{z}_i$. 
The data are summarized by $\{ \bm{n}_i, \bm{z}_i, \bm{x}_i(s) \}_{i=1}^m$. 
Denote the parameter set by $\bm{\theta}=\{\bm{\eta}, \gamma, \bm{\beta}, \psi\}$, consisting of the zero-inflation parameters $\bm{\eta} = (\eta_1, \ldots, \eta_{d_1})^{\top}$, baseline intensity parameter $\gamma$, intensity coefficients $\bm{\beta}=(\beta_1, \ldots, \beta_{d_2})^{\top}$, and storm-specific random effect parameter $\psi$. 

Let $L_i(\bm{\theta}| \bm{n}_i, \bm{z}_i, \bm{x}_i(s))$ denote the likelihood of storm $i$ based on the joint probability mass function (\ref{eq:ZInegmult}). 
For brevity, let $\bm{\mu}_i =\left(\mu_i(B_{i1}), \ldots, \mu_i(B_{i k_i}) \right)^{\top}$ denote the vector of spatial integrals $\mu_i(B_{ij}) = \int_{B_{ij}} \gamma \exp\{ \bm{x}_i(s)^{\top}\bm{\beta} \} ds $, $j=1,\ldots, k_i$, which are functions of the parameters $\gamma$ and $\bm{\beta}$. As mentioned earlier, we let $p_i = \left( 1 + \exp\{ -\bm{z}_i^{\top} \bm{\eta} \} \right)^{-1}$, which are functions of the parameters $\bm{\eta}$. 
The contribution of storm $i$ to the log-likelihood is 
\begin{align}
	&\ell_i (\bm{\theta}| \bm{n}_i, \bm{z}_i, \bm{x}_i(s))  = \log L_i(\bm{\theta}| \bm{n}_i, \bm{z}_i, \bm{x}_i(s)) \nonumber \\
	&\quad = I(  n_{i+} = 0 ) \log\left[ p_i + (1-p_i)\left( \frac{\psi^{-1}}{\psi^{-1} + \mu_{i+}} \right)^{\psi^{-1}} \right] +  \nonumber \\
	&\quad \qquad  I( n_{i+} > 0 ) \left\{  \log(1-p_i) + \log \Gamma(\psi^{-1}+n_{i+}) - \log \Gamma(\psi^{-1}) - \sum_{j=1}^{k_i} \log n_{ij}! \, + \right. \nonumber \\
	&\quad \qquad \quad \left. \psi^{-1} \left(\log \psi^{-1} - \log(\psi^{-1}+\mu_{i+}) \right) + 
	 \sum_{j=1}^{k_i} n_{ij}\left( \log \mu_i(B_{ij}) - \log(\psi^{-1}+\mu_{i+}) \right) \right\}, 
\end{align}
where 
$n_{i+} = \bm{1}^{\top}\bm{n}_i$ and $\mu_{i+} = \bm{1}^{\top}\bm{\mu}_i$. 
Then the log-likelihood function is
\begin{align} \label{eq:DiscreteZILikelihood}
	\ell(\bm{\theta}| \bm{n}, \bm{z}, \bm{x}(s) ) = 
	\sum_{i=1}^m \ell_i \left(\bm{\theta}| \bm{n}_i, \bm{z}_i, \bm{x}_i(s) \right).
\end{align}  
As is often the case with zero-inflated models, directly maximizing the log-likelihood (\ref{eq:DiscreteZILikelihood}) can prove challenging for distinguishing between the joint zeros from the excess zero and the grouped spatial point process components \citep[e.g.,][]{Hall2000}. We address this challenge with a tailored expectation-maximization (EM) algorithm for the zero-inflated grouped spatial point process. 
Let $\pi_i$ denote a latent indicator variable such that $\pi_i=1$ if $\bm{n}_i$ is generated by the excess zero component with probability $p_i$,  
and $\pi_i=0$ if $\bm{n}_i$ is generated from the grouped spatial point process with probability $1-p_i$. The log-likelihood with complete data $\{\bm{n}_i, \bm{z}_i, \bm{x}_i(s),  \pi_i\}_{i=1}^m$ is 
\begin{equation}
\begin{aligned}\label{eq:CompleteLogLikelihood}
	\ell_c(\bm{\theta}|\bm{n}, \bm{z}, \bm{x}(s), \bm{\pi}) &= \sum_{i=1}^{m} \pi_i\log\left[p_i(\bm{\theta}| \bm{z}_i) \right] + \\ 
	&\qquad \qquad (1-\pi_i)\log\left[(1-p_i( \bm{\theta}| \bm{z}_i ) )f_{i}(\bm{\theta}|\bm{n}_i,  \bm{x}_i(s))\right],
\end{aligned}
\end{equation}
where
\begin{align}
	f_i(\bm{\theta}|\bm{n}_i, \bm{x}_i(s)) =  \dfrac{\Gamma(\psi^{-1} + n_{i+})}{\Gamma(\psi^{-1}) \prod_{j=1}^{k_i} n_{ij}!}\ \left(\dfrac{\psi^{-1}}{\psi^{-1} + \mu_{i+}} \right)^{\psi^{-1}} \prod_{j=1}^{k_i} \left( \dfrac{\mu_i(B_{ij})}{ \psi^{-1} + \mu_{i+} } \right)^{n_{ij}},
\end{align}
is the probability mass function of the grouped spatial point process component. 
We initialize $\bm{\theta}^{(0)}$ by setting the zero-inflation coefficients $\bm{\eta}^{(0)}$ based on a storm-level logistic regression for whether a storm had zero claims. Similarly, using storm-subregions with positive claim counts, we initialize intensity coefficients $\bm{\beta}^{(0)}$ and baseline intensity $\gamma^{(0)}$ based on a zero-truncated negative binomial regression with a log link, where predictors are averaged within each storm-subregion. 
Lastly, we initialize the storm random effect parameter $\psi^{(0)}$ with a moment-based estimate from the zero-inflated negative binomial distribution. 
Using $\bm{\eta}^{(0)}$, we compute the initial fitted zero-inflation probabilities $p_i^{(0)} = \left( 1 + \exp \{ -\bm{z}_i^{\top} \bm{\eta}^{(0)} \} \right)^{-1}$, $i=1,\ldots,m,$ and their average $\bar{p}^{(0)} = \frac{1}{m}\sum_{i=1}^m p_i^{(0)}$, and then compute
\begin{align*}
	\psi^{(0)} = \frac{(s_n^2-\bar{n})(1-\bar{p}^{(0)})}{\bar{n}^2}-\bar{p}^{(0)},
\end{align*}
where $\bar{n}$ and $s_n^2$ are the sample mean and sample variance of the total number of claims from a storm $n_{i+}$, $i=1,\ldots, m$. 
For iteration $t$, the E-step sets the latent variables $\bm{\pi}^{(t)} =(\pi_1^{(t)}, \ldots, \pi_m^{(t)})^{\top}$ equal to their conditional expectations given $\bm{\theta}^{(t)}$. For $i=1,\ldots,m$,
\begin{align}
	\pi_i^{(t)} &= E(\pi_i|\bm{n}_i, \bm{z}_i, \bm{x}_i(s), \bm{\theta}^{(t)}) \nonumber \\
	&= \begin{cases}
		\dfrac{ p_i^{(t)} }{  p_i^{(t)} + \left(1-p_i^{(t)}\right)\left(\frac{ {\psi^{(t)}}^{-1} }{ {\psi^{(t)}}^{-1} + \mu_{i+}^{(t)}}\right)^{ {\psi^{(t)}}^{-1} }} \, , & n_{i+}=0, \\
		0, & n_{i+} > 0.
	\end{cases} \label{eq:Estep}
\end{align}
Equation \eqref{eq:Estep} shows that the conditional expectation of the latent zero-inflation indicator is only non-trivial when the observed storm-level claim vector is jointly zero. If $n_{i+}>0$, the observation cannot have arisen from the excess zero component, so $\pi_i^{(t)}=0$. If $n_{i+}=0$, the observed joint zero claim counts may have arisen from either the excess zero component or from the grouped spatial point process component generating zero claims, and the E-step computes the posterior probability of the former. 

The M-step updates estimates $\bm{\theta}^{(t+1)}$ by maximizing \eqref{eq:CompleteLogLikelihood} given the latent variables $\bm{\pi}^{(t)}$:
\begin{equation}\label{eq:Mstep}
	\bm{\theta}^{(t+1)}= \argmax_{\bm{\theta}} \ell_c (\bm{\theta}|\bm{n},\bm{z}, \bm{x}(s), \bm{\pi}^{(t)}).
\end{equation}
We repeat the E-step and M-step until convergence in the log-likelihood within some relative tolerance $\varepsilon_{tol}$, or until a maximum number of iterations $t_{max}$ is reached. The full EM algorithm procedure is detailed in Algorithm \ref{alg:EM_nll_convergence}. 
Under suitable conditions, the maximum likelihood estimator $\hat{\bm{\theta}}$ obtained from the EM algorithm is a consistent estimator and is asymptotically normally distributed \citep{Dempster1977}, with mean equal to the true parameters $\bm{\theta}$, and covariance $I(\bm{\theta})^{-1}$. 
Following standard likelihood-based inference, we estimate standard errors using the observed information matrix, which is computed as the negative Hessian of the observed data log-likelihood \eqref{eq:DiscreteZILikelihood} evaluated at the converged estimates \citep{cox1979theoretical}.

\begin{algorithm}[h!]
	\singlespacing
	\SetAlgoLined
	\SetKwInOut{Input}{Input}
	\SetKwInOut{Output}{Output}
	\Input{Data $\bm{n}, \bm{z}, \bm{x}(s)$, windows $\bm{W}$}
	\Output{MLE $\hat{\boldsymbol{\theta}}$}
	\BlankLine
	
	Set relative tolerance $\varepsilon_{tol} = 10^{-9}$\ and maximum iterations $t_{max}=500$\;
	Set iteration counter $t = 0$\;
	Initialize $\bm{\eta}^{(0)}$ by logistic regression on $\{I(n_{i+}=0)\}_{i=1}^{m}$ with predictors $\{\bm{z}_i\}_{i=1}^m$\; 
	Initialize $\bm{\beta}^{(0)}$ and $\gamma^{(0)}$ by zero-truncated negative binomial regression on $\{n_{ij}\}_{\{i,j : n_{ij}>0\}}$ with a log link on storm-subregion average predictors $\{\bar{x}_{ij}\}_{\{i,j : n_{ij}>0\}}$ \;
	Calculate $p_i^{(0)}=(1+\exp\{-\bm{z}_i^{\top}\bm{\eta}^{(0)} \})^{-1}$, $i=1,\ldots,m$\;
	Initialize $\psi^{(0)} = \frac{(s_n^2-\bar{n})(1-\bar{p}^{(0)})}{\bar{n}^2}-\bar{p}^{(0)}$ for sample mean $\bar{n}$ and sample variance $s_n^2$ \;

	Calculate initial $\bm{\pi}^{(0)}$ using equation (\ref{eq:Estep})\;
	Calculate initial $L^{(0)} = \ell_c(\bm{\theta}^{(0)}|\bm{n},\bm{z},\bm{x}(s),\bm{\pi}^{(0)})$ using equation (\ref{eq:CompleteLogLikelihood})\;
	\While{ $t \le t_{max}$  }{ 
		Update $\bm{\theta}^{(t+1)}$ using equation (\ref{eq:Mstep})\tcc*{M-step} 
		Calculate $\bm{\pi}^{(t+1)}$ using equation (\ref{eq:Estep}) \tcc*{E-step}  
		Calculate $L^{(t+1)} = \ell_c(\bm{\theta}^{(t+1)}|\bm{n},\bm{z},\bm{x}(s),\bm{\pi}^{(t+1)})$ using equation (\ref{eq:CompleteLogLikelihood})\;
		
		\If{
			$|L^{(t+1)} - L^{(t)}|/|L^{(t)}| < \varepsilon_{tol}$
		}{
			\textbf{break}\;
		}
		
		Set $t = t + 1$\;
	}
	\Return{$\bm{\theta}^{(t)}$}
	\BlankLine
	
	\caption{EM algorithm for zero-inflated spatial point process with grouped outcomes.}
	\label{alg:EM_nll_convergence}
\end{algorithm}

Lastly, the evaluation of (\ref{eq:DiscreteZILikelihood}) and (\ref{eq:CompleteLogLikelihood}) requires computing the two-dimensional spatial integrals $\mu_i(B_{ij})$ over each subregion $B_{ij}, j=1\ldots,k_i,$ in each storm $i=1,\ldots,m$. We employ the numerical quadrature originally proposed by \citet{berman1992approximating} for the estimation of inhomogeneous Poisson spatial point processes, where the integral $\mu_i(B_{ij})$ is approximated as a finite weighted sum of the integrand evaluated at $q_{ij}$ quadrature points $u_{ij,1}, \ldots, u_{ij,q_{ij}}$ within $B_{ij}$, with corresponding quadrature weights $w_{ij,1}, \ldots, w_{ij,q_{ij}}$: 
\begin{align}
	\mu_i(B_{ij}) &= \int_{B_{ij}} \gamma \exp\{\bm{x}_i(s)^{\top}\bm{\beta} \} ds \nonumber \\ 
	&\approx  \sum_{l=1}^{q_{ij}} \gamma \exp\{\bm{x}_i(u_{ij,l})^{\top}\bm{\beta} \} w_{ij,l}.
\end{align}
The quadrature points consist of locations where the intensity can be evaluated, which facilitates the incorporation of the more granular, densely measured exposure information within the regions. The quadrature weights sum to the area $|B_{ij}| = \int_{B_{ij}} 1 ds$, computed by dividing the region $B_{ij}$ into tiles, with the weight on each quadrature point equal to the area of its tile scaled by the number of quadrature points in that tile. We employ the more computationally efficient tiles based on a regular grid \citep{baddeley2000practical}; tiles based on the Dirichlet tessellation are another common option \citep{berman1992approximating}.

\section{Numerical studies}\label{sec:NumericalExperiment}

We conduct simulation exercises to illustrate the proposed estimation procedure and examine the finite-sample performance and asymptotic properties, as well as the effect of differing extents of spatial aggregation in the grouped loss outcome variable. 
We consider $m \in \{500,\, 1{,}000\}$ storms. For each storm $i$, the underlying generating process of the grouped loss outcomes $\bm{n}_i$ is the zero-inflated mixed Poisson spatial point process. The spatial point patterns of claim locations are generated within the unit square window $W = [0,1] \times [0,1]$. Excess zero probabilities are governed by a storm-level predictor $z_i$ generated from a standard normal distribution, and incorporated through a logit link so that $p_i = (1 + \exp\{ -(\eta_0 + z_i \eta_1)\})^{-1}$. Conditional on sampling from the spatial point process, the storm-specific random effect $u_i$ is drawn from a gamma distribution with scale parameter $\psi$ and shape parameter $\psi^{-1}$. Then the conditional intensity is
\begin{align*}
	\lambda_i(s|u_i) = u_i \gamma \exp\{ x_{i,1}(s) \beta_1 + x_{i,2}(s) \beta_2 \}, \qquad s \in W,
\end{align*} 
where predictors $\bm{x}_i(s) = (x_{i,1}(s), x_{i,2}(s))^{\top}$ include a spatially varying predictor $x_{i,1}(s) = (|s - s_i| + 0.1)^{-1}$ that decays by distance from a spatially uniformly drawn storm-specific location $s_i \in W$, and a spatially invariant predictor $x_{i,2}(s) = x_{i,2}$ sampled independently from a standard uniform distribution. Finally, the simulated zero-inflated spatial point pattern of claims is grouped into multivariate claim counts $\bm{n}_i=(n_{i1}, \ldots, n_{ik})$ within $k$ partitioned subregions of $W$. The full simulation algorithm is provided in the \hyperref[sec:Appendix]{Appendix}. 

\begin{figure}[h!]
	\centering
	\includegraphics[width=0.45\linewidth]{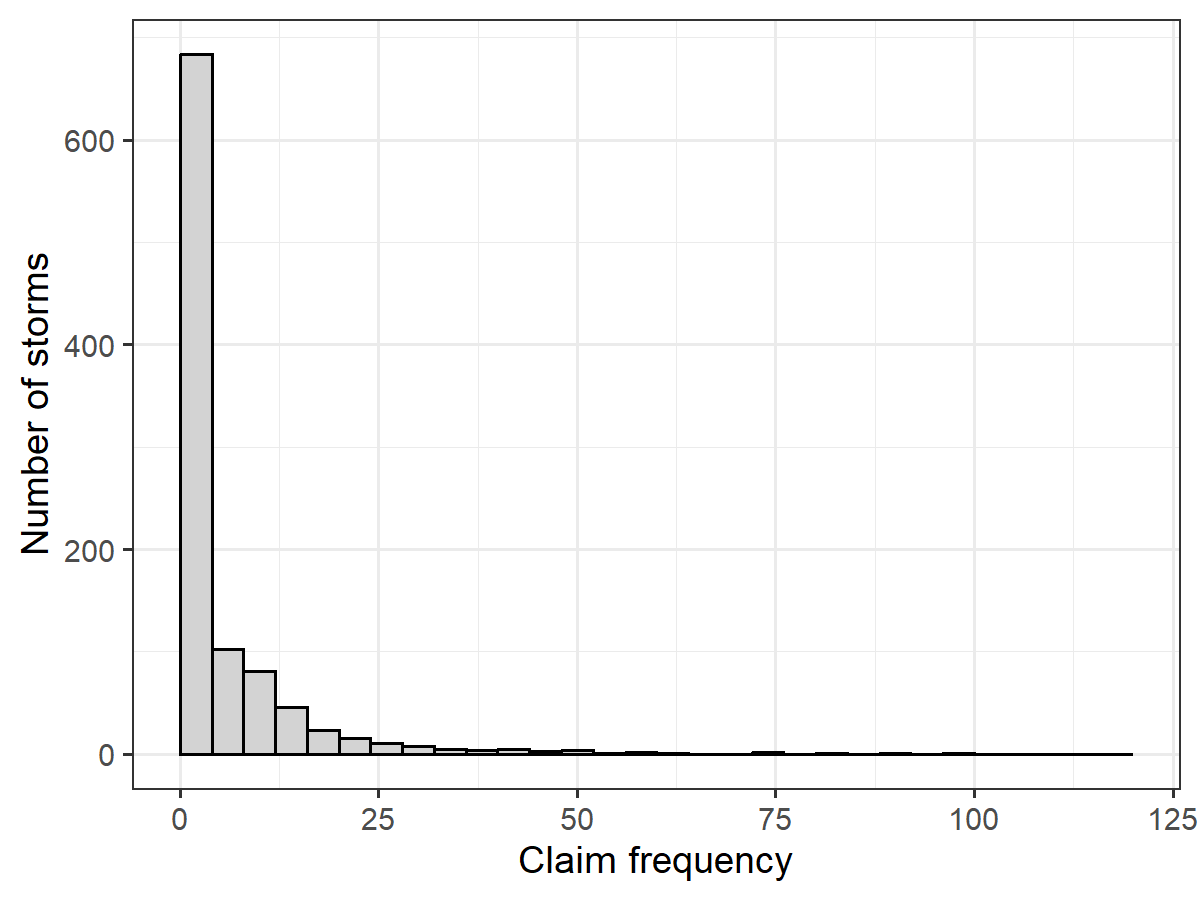}
	\caption{Distribution of storm-level claim frequency for $m=1{,}000$ storms from one simulation replication.}
	\label{fig:sim_NiDensity}
\end{figure}

True values for the parameter vector $\bm{\theta}=(\eta_0, \eta_1, \beta_1, \beta_2, \gamma, \psi)^{\top}$ are given in Table \ref{tab:ParameterEstimates}, chosen to reflect characteristics of the storm setting such as excess zeros and substantial storm-specific heterogeneity. 
Figure \ref{fig:sim_NiDensity} displays a histogram of the storm-level claim frequency for $m=1{,}000$ storms from one simulation replication, indicating the zero-inflation. On average across the simulation replications, 45\% of the simulated storms have zero claims. 

\begin{figure}[h!]
	\centering 
	\includegraphics[width=0.45\linewidth]{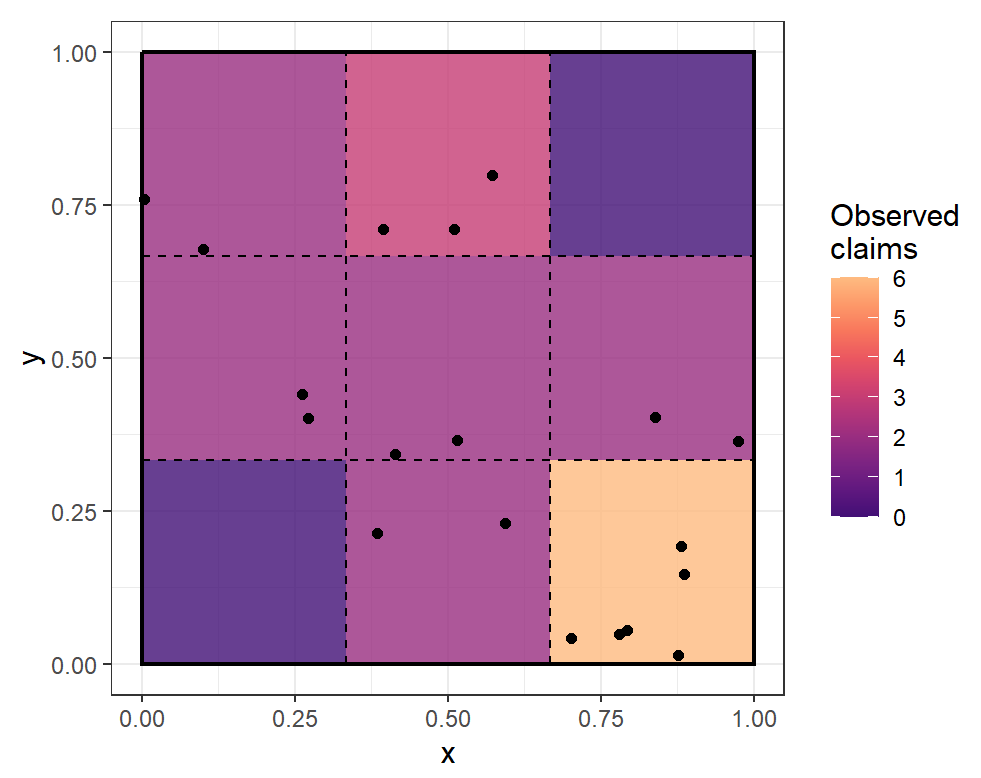}
	\caption{Example of an underlying simulated spatial point pattern, with observed claim counts spatially grouped into $3\times 3$ subregions.}
	\label{fig:Discretized3x3}
\end{figure}

In particular, we consider different extents of spatial aggregation of the grouped loss outcomes, and divide $W$ into $b \times b$ grid-based subregions, for $b \in \{2, 3, 4, 5\}$. As an example, Figure \ref{fig:Discretized3x3} displays one underlying simulated point pattern of claim locations from a storm, spatially grouped into $3\times 3$ subregions, producing 9 observed subregional claim counts. We estimate the model using the EM algorithm procedure in Algorithm \ref{alg:EM_nll_convergence}, with $50 \times 50$ quadrature points in the spatial integrals. 
As a full information reference for the grouped loss outcomes, we also consider a fully granular case where the simulated point pattern of claim locations, as well as the granular predictors, are exactly observed without spatial grouping. This reference can be interpreted as the limiting case as the $b \times b$ partition becomes arbitrarily fine, and corresponds to the replicated spatial point patterns modeled in \citet{Gao2022}. Due to the zero-inflation, we also estimate this fully granular case using the EM algorithm, where the grouped spatial point process component of the log-likelihood is replaced by the spatially continuous version given in Section 4.1 of \citet{Gao2022}.

Table \ref{tab:ParameterEstimates} reports the estimation results based on 200 simulation replications, each using $m=500$ or $m=1{,}000$ storms, with loss outcomes grouped at the $3\times 3$ and $5\times 5$ subregional levels, along with the fully granular exact claim location reference case. The results for the $2 \times 2$ and $4 \times 4$ cases are qualitatively similar, so are omitted for brevity. We report the average relative bias and empirical standard deviation of the estimates, along with the average standard errors and coverage probabilities of the 95\% confidence intervals. The relative bias of the estimates is low and the average standard errors are close to the empirical standard deviations, indicating that the estimated standard errors effectively reflect the sampling variability of the parameter estimates. 
For both the coarser $3\times 3$ and the finer $5\times 5$ subregions of grouped losses, as well as for the fully granular case, as the sample size increases from $m=500$ to $m=1{,}000$, the relative bias decreases and the empirical standard deviations and average standard errors decrease by roughly $1/\sqrt{2}$, supporting $\sqrt{m}$-consistency. The coverage probabilities of the 95\% confidence intervals also improve with increasing sample size.

\begin{table}[htbp]
	\centering
	\caption{Average relative bias, empirical standard deviation, average standard error, and coverage probabilities of parameter estimates for the zero-inflated mixed Poisson point process model based on 200 simulation replications.}
	{\footnotesize 
		\begin{tabular}{r l r r r r r}
			\toprule
			\multicolumn{1}{c}{$m$} &       & \textbf{True} & \multicolumn{4}{c}{\textbf{3x3 subregions}} \\
			\cmidrule{4-7}
			&       &       & \textbf{RelBias} & \textbf{EmpSD} & \textbf{AvgSE} & \textbf{Coverage} \\
			\midrule
			500   & $\eta_0$ & -1.4  & 0.0684 & 0.5695 & 0.5150 & 0.91 \\
			& $\eta_1$ & 1     & 0.0605 & 0.3519 & 0.3373 & 0.93 \\
			& $\beta_1$ & 0.5   & -0.0099 & 0.0122 & 0.0115 & 0.91 \\
			& $\beta_2$ & -1    & 0.0217 & 0.2701 & 0.2703 & 0.95 \\
			& $\gamma$ & 3     & 0.0349 & 0.4927 & 0.5268 & 0.97 \\
			& $\psi$ & 2     & 0.0032 & 0.3441 & 0.3439 & 0.92 \\
			\midrule
			$1{,}000$  & $\eta_0$ & -1.4  & 0.0406 & 0.3501 & 0.3333 & 0.96 \\
			& $\eta_1$ & 1     & 0.0215 & 0.2308 & 0.2195 & 0.94 \\
			& $\beta_1$ & 0.5   & -0.0085 & 0.0081 & 0.0081 & 0.91 \\
			& $\beta_2$ & -1    & 0.0134 & 0.1723 & 0.1898 & 0.98 \\
			& $\gamma$ & 3     & 0.0245 & 0.3385 & 0.3674 & 0.98 \\
			& $\psi$ & 2     & -0.0015 & 0.2454 & 0.2428 & 0.93 \\
			\midrule
			
			\multicolumn{1}{c}{$m$} &       & \textbf{True} & \multicolumn{4}{c}{\textbf{5x5 subregions}} \\
			\cmidrule{4-7}
			&       &       & \textbf{RelBias} & \textbf{EmpSD} & \textbf{AvgSE} & \textbf{Coverage} \\
			\midrule
			500   & $\eta_0$ & -1.4  & 0.0685 & 0.5759 & 0.5151 & 0.91 \\
			& $\eta_1$ & 1     & 0.0605 & 0.3562 & 0.3374 & 0.93 \\
			& $\beta_1$ & 0.5   & -0.0081 & 0.0117 & 0.0104 & 0.92 \\
			& $\beta_2$ & -1    & 0.0217 & 0.2704 & 0.2703 & 0.95 \\
			& $\gamma$ & 3     & 0.0315 & 0.4890 & 0.5228 & 0.97 \\
			& $\psi$ & 2     & 0.0033 & 0.3432 & 0.3440 & 0.92 \\
			\midrule
			$1{,}000$  & $\eta_0$ & -1.4  & 0.0407 & 0.3535 & 0.3333 & 0.96 \\
			& $\eta_1$ & 1     & 0.0215 & 0.2311 & 0.2195 & 0.94 \\
			& $\beta_1$ & 0.5   & -0.0060 & 0.0082 & 0.0073 & 0.90 \\
			& $\beta_2$ & -1    & 0.0134 & 0.1724 & 0.1898 & 0.98 \\
			& $\gamma$ & 3     & 0.0204 & 0.3419 & 0.3643 & 0.98 \\
			& $\psi$ & 2     & -0.0015 & 0.2447 & 0.2429 & 0.93 \\
			\midrule

			\multicolumn{1}{c}{$m$} 
			&       
			& \textbf{True} 
			& \multicolumn{4}{c}{\textbf{Fully granular}} \\
			\cmidrule{4-7}
			&       
			&       
			& \textbf{RelBias} 
			& \textbf{EmpSD} 
			& \textbf{AvgSE} 
			& \textbf{Coverage} \\
			\midrule
			500   
			& $\eta_0$ 
			& -1.4  
			& 0.0684 
			& 0.5694 
			& 0.5151 
			& 0.91 \\
			
			& $\eta_1$ 
			& 1     
			& 0.0605 
			& 0.3518 
			& 0.3374 
			& 0.93 \\
			
			& $\beta_1$ 
			& 0.5   
			& -0.0052 
			& 0.0100 
			& 0.0094 
			& 0.92 \\
			
			& $\beta_2$ 
			& -1
			& 0.0217
			& 0.2701 
			& 0.2703 
			& 0.95 \\
			
			& $\gamma$
			& 3     
			& 0.0268 
			& 0.4788 
			& 0.5186 
			& 0.97 \\
			
			& $\psi$ 
			& 2     
			& 0.0033 
			& 0.3441 
			& 0.3439 
			& 0.92 \\
			\midrule
			
			$1{,}000$ 
			& $\eta_0$
			& -1.4  
			& 0.0406
			& 0.3499 
			& 0.3333 
			& 0.96 \\
			
			& $\eta_1$ 
			& 1     
			& 0.0214
			& 0.2307 
			& 0.2195 
			& 0.94 \\
			
			& $\beta_1$ 
			& 0.5   
			& -0.0034
			& 0.0067 
			& 0.0066 
			& 0.94 \\
			
			& $\beta_2$ 
			& -1   
			& 0.0134 
			& 0.1723 
			& 0.1898 
			& 0.98 \\
			
			& $\gamma$
			& 3     
			& 0.0160
			& 0.3332 
			& 0.3615 
			& 0.98 \\
			
			& $\psi$
			& 2 
			& -0.0016
			& 0.2454 
			& 0.2428 
			& 0.93 \\
			\bottomrule
		\end{tabular}%
	}
	\label{tab:ParameterEstimates}
\end{table}%

Table \ref{tab:ParameterEstimates} shows that the estimates behave comparably for the same underlying data generating process relative to the fully granular case, even when the observed loss outcomes are grouped at different levels of spatial aggregation. The metrics are similar between the $3 \times 3$, $5 \times 5$, and fully granular cases at both sample sizes. 
For the parameters $\beta_1$ and $\gamma$ most directly related to the within-storm spatial heterogeneity, the bias and standard errors decrease slightly as the granularity of the spatial aggregation increases. However, the differences between the grouped and fully granular cases are very small, and the remaining parameters, including those for zero-inflation and storm-level heterogeneity, exhibit nearly identical performance across the $3\times 3$, $5\times 5$, and fully granular cases. 
Thus, the proposed method for grouped loss outcomes produces finite-sample estimation results that remain comparable to the fully granular observation setting of exact claim locations.

\begin{figure}[H]
	\centering
	\includegraphics[width=1\linewidth]{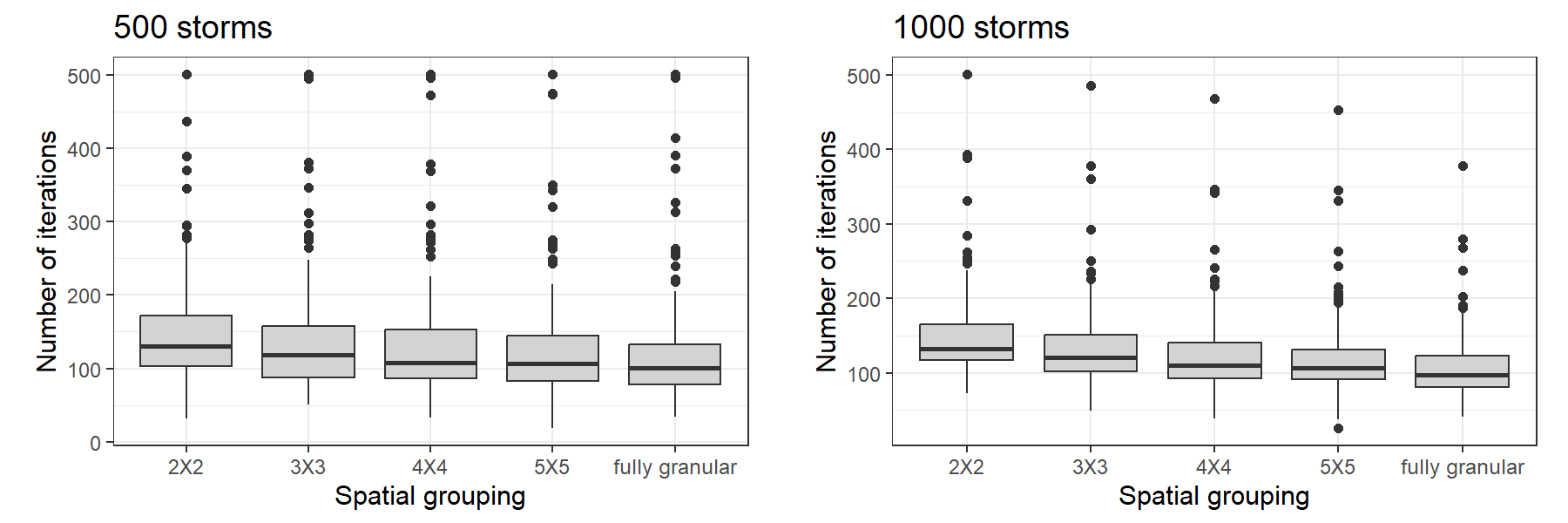}
	\caption{Boxplots of the number of EM algorithm iterations until convergence under various levels of spatial aggregation of the loss outcomes, using 500 storms (left) and $1{,}000$ storms (right), based on 200 simulation replications.}
	\label{fig:Boxplots}
\end{figure}

On the computational end, we note that the EM algorithm takes longer to converge when the outcomes are spatially grouped at coarser levels. Figure \ref{fig:Boxplots} depicts boxplots of the number of EM algorithm iterations until convergence over 200 simulation replications, under different levels of spatial aggregation of the observed loss outcomes, for $m=500$ storms (left) and $m=1{,}000$ storms (right). Finer levels of spatial aggregation require a lower number of EM iterations to converge, with the fully granular case representing the finest observation setting. In addition, the variability in the number of iterations across experiments decreases as the sample size increases. 

The numerical experiments show that the proposed spatial point process-derived method can be effectively estimated using spatially grouped loss outcomes with more densely measured predictors, across various levels of aggregation for the observed outcome.

\section{Storm loss data}\label{sec:Data}

\begin{figure}[htbp]
	\centering
	\includegraphics[width=0.64\textwidth]{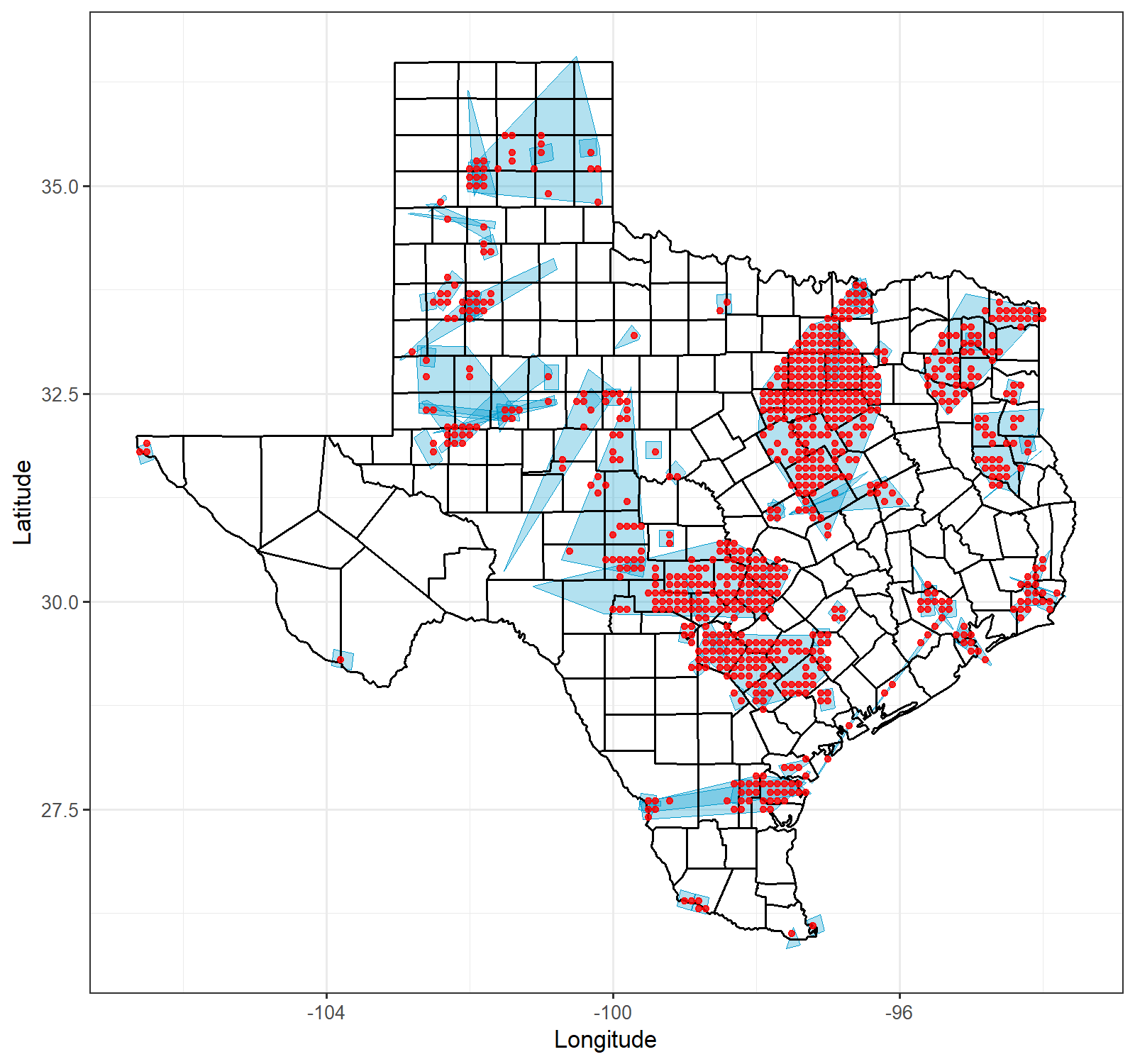}
	\caption{Storm windows for 2023 storms in Texas as blue polygons, with exposure locations marked by red points.}
	\label{fig:stormWindows2023}
\end{figure}

To demonstrate our proposed method on real storm loss data, we construct a replicated spatial dataset of property losses arising from a series of storm events that are linked to granular weather and property exposure information. 
Our sample comprises storm events in Texas from 2009 to 2023, obtained from the NOAA Storm Events Database maintained by the National Centers for Environmental Information \citepalias{NOAAstormdata}.
We extract flash flood and flood storm events, defined respectively as a rapid rise of water into a normally dry area (for example, due to intense rainfall), and an inundation of water in a normally dry area. 
Flash flooding may transition into flooding, yet minor flooding from heavy rain is considered flooding to distinguish its non-life-threatening nature from flash floods. Because of the substantial overlap in definitions and reliance on professional discretion in their coding \citepalias{NOAAstormdataprep}, we include both event types in our sample. 
In particular, we exclude coastal and lakeshore flooding events associated with a vertical rise in water level from wind or high tide, as well as hurricane-related events, to focus on pluvial-related storm events with a defined storm path. 
We recognize separate storms by the \textit{episode}, which identifies the entire meteorological storm system, and may consist of multiple smaller events (e.g., flood, flash flood, or other perils). Based on the beginning and ending points of the event paths within a storm episode, we obtain the spatial window for each storm as the convex hull of the path locations. In total, we observe 1,332 storms and their associated dates and spatial windows, a subset of which are depicted as blue polygons in Figure \ref{fig:stormWindows2023} for the year 2023. 

We link each storm to granular property exposure information by overlaying the storm spatial window on top of the in-force policies from the US National Flood Insurance Program (NFIP) to identify the insured properties affected by the storm \citep{NFIPpolicydata}. 
The publicly-run NFIP has been the primary provider of residential flood insurance in the US for the past 50 years, although the data are only representative of the in-force policy base from 2009 onwards, hence the start of our sample. The program is responsible for more than 95\% of residential flood policies sold, with a small and growing private residential flood market for excess policies \citep{kousky2018emerging}. 
Policies cover physical damage directly caused by a flood, such as water entering the property from the ground up due to heavy rainfall, storm surge, or river overflow \citepalias{NFIPclaimshandbook}. 
For each storm, we consider the standard residential policies for single-family residences that are in-force during the storm, and whose location falls within the spatial window of the storm. 
For example, the exposure locations for the storms from 2023 are marked by red points in Figure \ref{fig:stormWindows2023}. 
While insurers may have access to proprietary geocoded property exposure information, the property locations in the public NFIP data are less precise and rounded to the nearest 0.1 degrees longitude and latitude. For the purposes of demonstrating our proposed methodology, where individual properties would not share the exact same location, we slightly spatially jitter the exposure locations by adding uniform $U(-0.05, 0.05)$ random noise to the coordinates. 

\begin{figure}[htbp]
	\centering
	\includegraphics[width=0.57\textwidth]{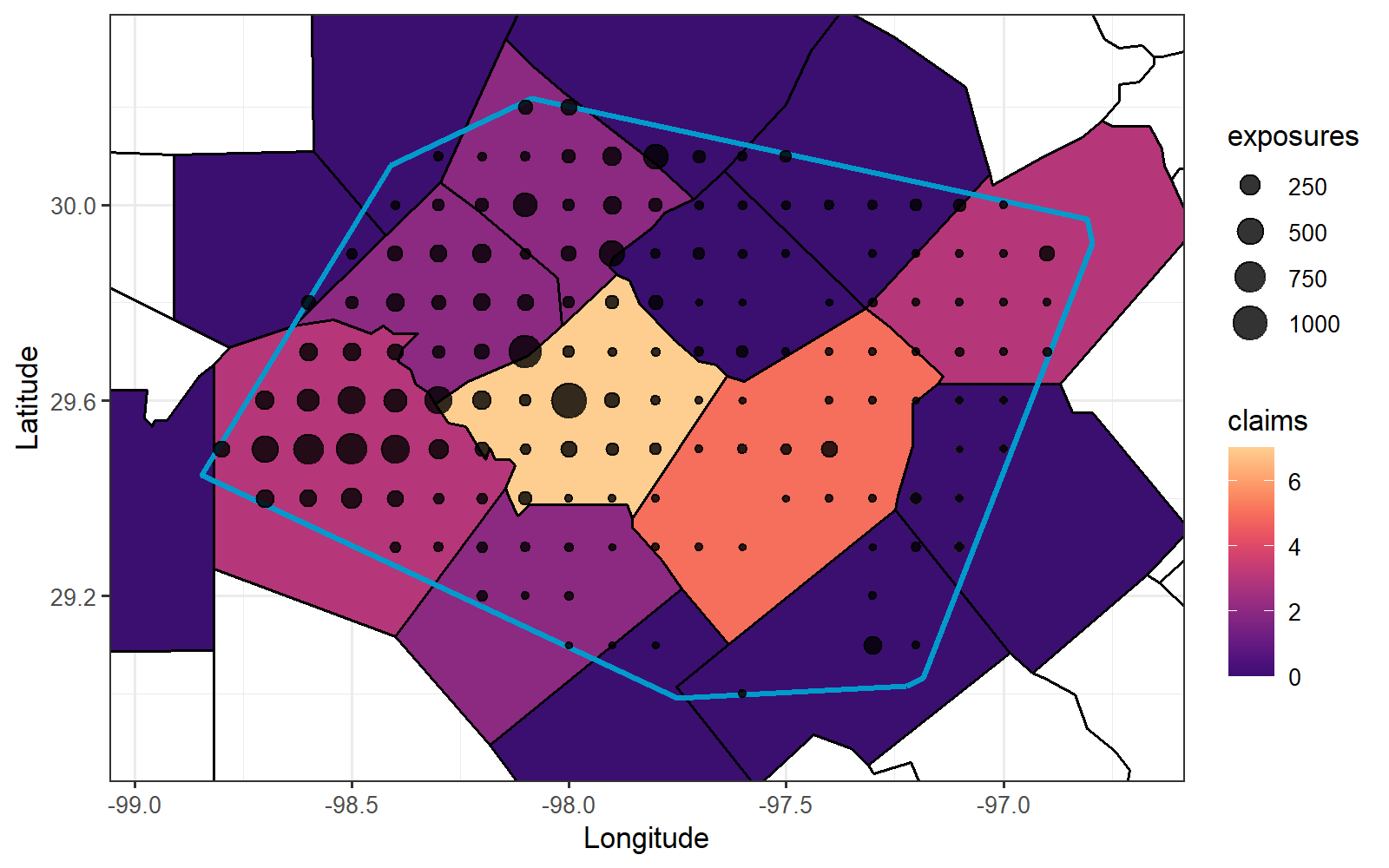}
	\caption{Example storm map with spatial window outlined in blue, exposure locations as black points, and counties color-coded by the number of observed claims.}
	\label{fig:mapOneStm} 
\end{figure}

For each storm, we similarly identify the associated claims based on the spatial window and date range of the storm. 
Specifically, a claim can be associated with a storm whose spatial storm window encompasses the claim location, and whose storm dates encompass the reported date of loss for the claim (i.e., when water first entered the property). We allow a plausible margin of one week on the storm dates in the claims matching to account for potential inaccuracies in the self-reported date of loss. In cases of multiple candidate storms, we attribute the claim to the storm with the closest start date to the reported date of loss. 
Under this approach, our replicated spatial data comprise 1,332 storms that collectively produced 7,337 claims. 
For each storm, we observe characteristics of the affected properties and associated claims, but FEMA does not provide an identifier that allows users to link claims and policies. 
Figure \ref{fig:mapOneStm} illustrates one example storm, with the spatial window outlined in blue, exposure locations indicated by black points, and counties color-coded by the number of claims.

While we make our best effort to link pluvial-related storm events to water property damage exposure and claims using publicly available data, we recognize the limitations in the construction of this storm loss dataset. 
The flood and flash flood NOAA Storm Events may not be exclusively caused by heavy rainfall from severe convective storms. 
In addition, severe thunderstorms and intense rainfall may lead to flash flood and flood damage covered under the NFIP, but may also produce wind damage and wind-driven rain that is excluded from the NFIP (for example, if the roof is damaged and water enters through the ceiling). 
Similarly, the flood and flash flood storm events may lead to water property damage, yet are not exhaustive causes of losses covered by the NFIP, which also includes fluvial flooding. 
Despite these constraints, the replicated storms offer an illustrative setting for us to showcase the applicability of our method in addressing localized storm heterogeneity. 

\begin{figure}[htbp]
	\centering
	\includegraphics[width=0.45\linewidth]{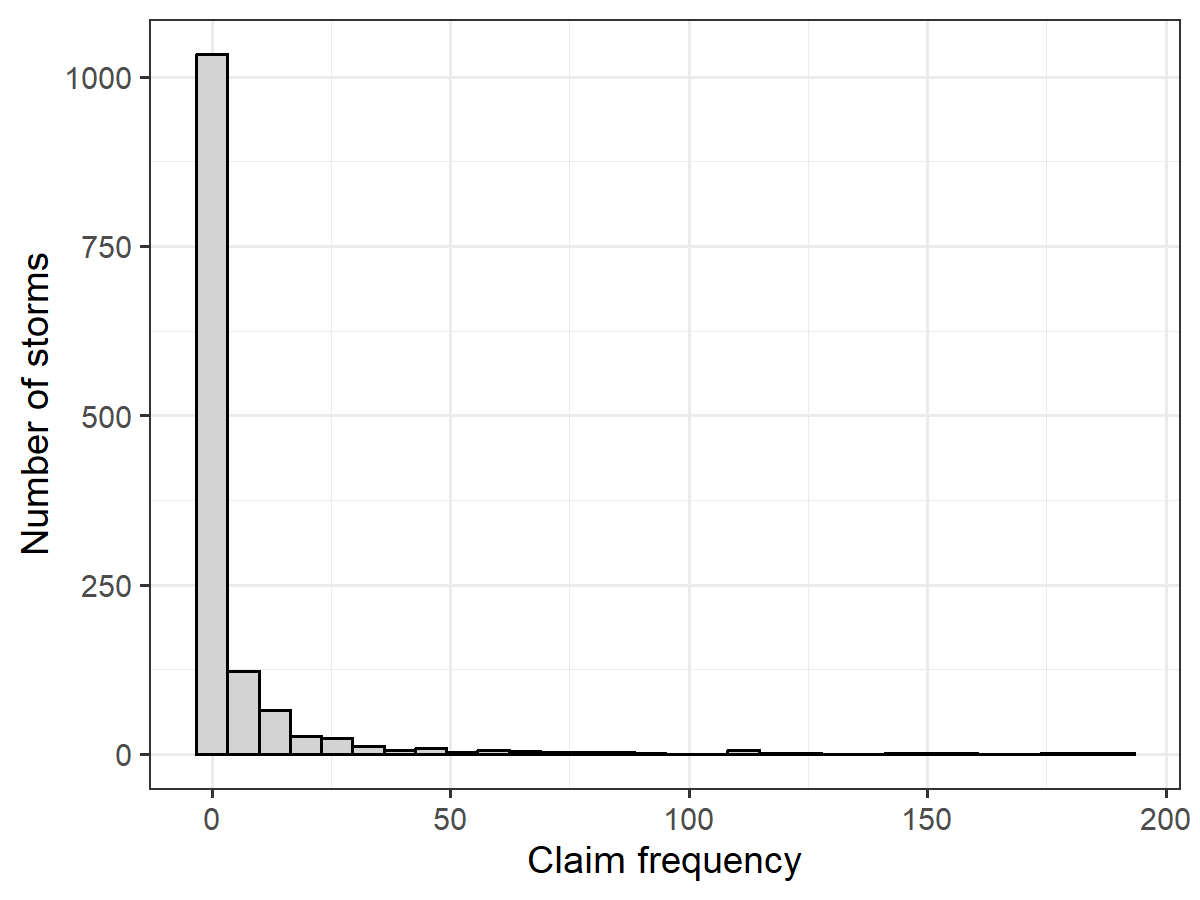}
	\caption{Distribution of storm-level claim frequency.}
	\label{fig:NiDensity}
\end{figure}

We highlight a few characteristics of the storm loss data. First, the claim count outcomes exhibit zero inflation, where 755 (approximately 57\%) of the 1,332 storms did not produce claims. Figure \ref{fig:NiDensity} shows the distribution of storm-level claim frequency, featuring a large probability mass at zero. 

Second, there is substantial heterogeneity in claim frequency both within and across storms. The replicated multivariate claim counts are unbalanced, where different storms may affect different numbers of subregions. We consider two levels of spatial aggregation for the grouped storm loss outcomes: coarser county-level claim counts (e.g., as portrayed in Figure \ref{fig:mapOneStm}); and more granular community-level claim counts grouped at the one decimal place of latitude and longitude. 
Table \ref{tab:ClaimStatsSubregion} provides descriptive statistics for the spatially grouped number of claims in a storm at both levels of aggregation, where each within-storm characteristic in a row is summarized across all of the storms. On average, a storm affects 4.2 counties, although there is substantial variation where some storms impact only a single county, and others affect as many as 42 counties. In addition, within a storm, there can be substantial variation in the number of claims across subregions, where the within-storm standard deviation is dampened by the joint zero-inflation. 

\begin{table}[htbp]
	\centering
	\singlespacing
	\caption{Descriptive statistics for the spatially grouped claim frequency in a storm at the county and community levels of aggregation.}
	\begin{tabular}{lrrrr}
		\toprule
		\multicolumn{1}{p{13em}}{\textbf{Within-storm characteristic}} & \multicolumn{4}{c}{\textbf{Across all storms}} \\
		\cmidrule{2-5}          & \multicolumn{1}{c}{\textbf{Mean}} & \multicolumn{1}{c}{\textbf{SD}} & \multicolumn{1}{c}{\textbf{Min}} & \multicolumn{1}{c}{\textbf{Max}} \\
		\midrule
		Counties in a storm &       &       &       &  \\
		\qquad Number of counties & 4.2   & 5.6   & 1     & 42 \\
		\qquad Avg claim frequency & 1.1   & 3.2   & 0     & 37 \\
		\qquad SD of claim frequency & 1.4   & 4.7   & 0     & 68.8 \\
		\midrule
		Communities in a storm &       &       &       &  \\
		\qquad Number of communities & 22.7  & 53.6  & 1     & 426 \\
		\qquad Avg claim frequency & 0.4   & 1.2   & 0     & 15.9 \\
		\qquad SD of claim frequency & 0.6   & 1.8   & 0     & 19.5 \\
		\bottomrule
	\end{tabular}%
	\label{tab:ClaimStatsSubregion}
\end{table}%

To account for the observed heterogeneity in storm losses, we incorporate granular property exposure and weather predictors. The property exposure characteristics obtained from the NFIP include the building age in years, flood zone designation, coverage amount, and whether the building is elevated above ground level with no basement. We group the NFIP flood zone designations into high risk (A and V, where flood insurance purchase is mandatory for a mortgage), moderate risk (B), and low risk (C). Since X is considered low-to-moderate and overlaps both B and C, we treat it as the baseline category, along with the undetermined risk zone D. 
The NFIP insures residential homes up to a limit of \$250,000, and we define a binary variable for whether the coverage amount is below the limit as an indication of the property value. 

In addition to property exposure information, we also consider spatially varying weather characteristics. For each storm, we obtain daily Texas weather station data from the Global Historical Climatology Network \citep{Menne2012} across 3,872 stations for precipitation and 132 stations for the fastest consecutive 2-minute average wind speed. 
Insurers often have access to higher-resolution maps or weather radar information, 
so for the purposes of demonstrating our method, we mimic the spatially continuous nature of high-resolution weather data by interpolating the weather station measurements on each day where a storm occurred. 
We apply inverse distance weighting, a widely-used spatial interpolation method that assigns values to unobserved locations based on a weighted average of nearby observations, where weights decrease with distance. This approach is effective for the irregularly spaced station networks, as it preserves local data characteristics while enabling smooth interpolation across space \citep{Panigrahi2021}.

Table \ref{tab:Covariates} presents a summary of property exposure and weather predictors, distinguishing between storms that resulted in claims and those that did not. 
On average, storms that produced claims are associated with a larger number of insured properties, higher proportion of non-elevated buildings, and lower proportion of policies insured below the coverage limit. Storms that produced claims are also associated with higher within-storm maximum precipitation and wind speed on average. 
Incorporating these densely measured exposure and weather-related predictors helps us to characterize the localized heterogeneity exhibited by the multivariate zero-inflated storm loss outcomes.

\begin{table}[H]
	\centering
	\singlespacing
	\caption{Descriptive statistics for property and weather characteristics by whether the storm produced claims.}
	\begin{tabular}{lrrrrr}
		\toprule
		\textbf{Within-storm characteristic} & \multicolumn{2}{c}{\textbf{Storms with no claims}} &       & \multicolumn{2}{c}{\textbf{Storms with claims}} \\
		\cmidrule{2-3}\cmidrule{5-6}          & \textbf{Mean} & \textbf{SD} &       & \textbf{Mean} & \textbf{SD} \\
		\midrule
		Number of exposures (000s) & 0.97  & 3.82  &       & 12.80 & 29.20 \\
		Average building age (years) & 35.31 & 12.58 &       & 33.28 & 7.52 \\
		Flood zone risk &       &       &       &       &  \\
		\qquad High (\%) & 49.73 & 29.91 &       & 41.14 & 22.33 \\
		\qquad Moderate (\%) & 2.20  & 8.62  &       & 3.42  & 8.83 \\
		\qquad Low (\%) & 10.19 & 20.02 &       & 8.22  & 13.83 \\
		\qquad Undetermined (\%) & 37.88 & 28.71 &       & 47.21 & 23.16 \\
		Coverage &       &       &       &       &  \\
		\qquad Under \$250k (\%) & 68.18 & 24.66 &       & 58.07 & 18.12 \\
		\qquad At least \$250k (\%) & 31.82 & 24.66 &       & 41.93 & 18.12 \\
		Elevated building &       &       &       &       &  \\
		\qquad Elevated (\%) & 11.67 & 16.33 &       & 9.27  & 9.86 \\
		\qquad Not elevated (\%) & 88.33 & 16.33 &       & 90.73 & 9.86 \\
		Precipitation (mm) &       &       &       &       &  \\
		\qquad Within-storm min & 9.48  & 14.56 &       & 8.35  & 14.80 \\
		\qquad Within-storm max & 25.34 & 32.17 &       & 65.91 & 56.38 \\
		Wind speed (m/s) &       &       &       &       &  \\
		\qquad Within-storm min & 10.07 & 3.16  &       & 8.96  & 2.92 \\
		\qquad Within-storm max & 11.25 & 3.60  &       & 12.40 & 3.51 \\
		\bottomrule
	\end{tabular}%
	\label{tab:Covariates}%
\end{table}%

\section{Application}\label{sec:Application}

We illustrate the zero-inflated mixed Poisson spatial point process model on the grouped storm loss outcomes, at both the county and the community levels of spatial aggregation. The model is estimated with the 2009-2019 storms using the EM procedure detailed in Algorithm \ref{alg:EM_nll_convergence}, and the 2020-2023 storms are left out for out-of-sample prediction. The granular spatially-varying property exposure and weather characteristics described in Section \ref{sec:Data} are incorporated in the point process, and 
we include aggregate storm-level characteristics to model the joint zero-inflation probabilities, such as the number of affected properties, area of the storm window, and maximum precipitation and wind speed within the window. The Berman-Turner spatial integral quadrature points consist of the property exposure locations within each storm. 

The grouped loss outcomes under our zero-inflated point process jointly follow a multivariate zero-inflated negative binomial distribution, with negative binomial component mean parameters obtained from the intensity measure. To demonstrate the value of incorporating more granular predictors in addressing localized storm heterogeneity, we compare our proposed method to a multivariate zero-inflated negative binomial regression, applied directly on the joint subregional claim counts $\bm{n}_i$ arising from a given storm $i$. 
Notably, the multivariate count regression comparison model also accounts for the dependence between subregional claim counts within a storm, overdispersion, and storm-level heterogeneity in the joint excess zeros, but instead uses spatially aggregated covariates that match the level of observation of the count outcomes (e.g., the average insured building age in each county), as in a typical rectangular dataset. 
The multivariate zero-inflated negative binomial (MZINB) regression can be viewed as a special case of our proposed zero-inflated mixed Poisson point process-derived method when the predictors and outcomes are all spatially grouped at the same level.

\subsection{Estimation results and fit}

Table \ref{tab:EstimatedCoefficients} presents the estimation results for the proposed zero-inflated mixed Poisson spatial point process for grouped losses at both the county and the community levels. At the storm level, larger storm areas and more affected properties are associated with higher probabilities of producing claims. Among the densely measured weather characteristics, higher precipitation and higher wind speed are both associated with higher claim intensity. Lastly, property characteristics such as higher age and not being elevated above ground are associated with higher claim intensity. Because the flood zone designations primarily reflect fluvial flood risk, they may not align intuitively with the estimated claim intensity in our application.

\begin{table}[h!]
	\centering
	\singlespacing
	\caption{Estimation results for the proposed zero-inflated mixed Poisson spatial point process for grouped losses at the county level and at the community level.}
	\resizebox{\textwidth}{!}{
		\begin{tabular}{l r r l r r l}
			\toprule
			\textbf{Parameter} 
			& \multicolumn{3}{c}{\textbf{County-level}} 
			& \multicolumn{3}{c}{\textbf{Community-level}} \\
			\cmidrule{2-4}\cmidrule{5-7}
			& \textbf{Est.} 
			& \textbf{Std.\@ error} 
			& 
			& \textbf{Est.} 
			& \textbf{Std.\@ error}
			&  \\
			\cmidrule{1-4}\cmidrule{5-7}
			
			Zero-inflation coefficients & & & & & & \\
			\qquad Intercept & 0.042 & 2.155 & & -0.512 & 2.159 & \\
			\qquad log number of exposures & -0.631 & 0.108 & *** & -0.615 & 0.108 & *** \\
			\qquad log area & -0.813 & 0.261 & ** & -0.834 & 0.265 & ** \\
			\qquad log max precipitation & 0.209 & 0.170 & & 0.184 & 0.169 & \\
			\qquad log max wind speed & -0.152 & 0.812 & & 0.032 & 0.805 & \\ 
			\addlinespace
			
			Point process coefficients & & & & & & \\
			\qquad log precipitation & 0.404 & 0.029 & *** & 0.356 & 0.024 & *** \\
			\qquad log wind speed & 0.599 & 0.134 & *** & 0.842 & 0.114 & *** \\
			\qquad log building age & 1.205 & 0.101 & *** & 1.058 & 0.046 & *** \\
			\qquad Flood zone risk & & & & & & \\
			\qquad \qquad Undetermined (reference) & & & & & & \\
			\qquad \qquad High & -0.589 & 0.139 & *** & -0.087 & 0.084 & \\
			\qquad \qquad Moderate & 0.790 & 0.184 & *** & 0.600 & 0.146 & *** \\
			\qquad \qquad Low & -0.139 & 0.145 & & 0.200 & 0.095 & * \\
			\qquad Coverage under \$250k & 2.142 & 0.519 & *** & -0.173 & 0.087 & * \\
			\qquad Elevated & -0.786 & 0.225 & *** & -0.076 & 0.109 & \\ 
			\addlinespace
			
			Point process parameters & & & & & & \\
			\qquad Baseline $\gamma$ & 0.028 & 0.016 &  & 0.147 & 0.047 & ** \\
			\qquad $\psi$ & 2.487 & 0.181 & *** & 2.451 & 0.178 & *** \\
			\bottomrule
			\multicolumn{7}{l}{\footnotesize*$p<0.05$, **$p<0.01$, ***$p <0.001$.} \\
		\end{tabular}
	}
	\label{tab:EstimatedCoefficients}
\end{table}

In general, the signs of the statistically significant coefficient estimates are consistent across the two spatial aggregation levels. The exception is the coefficient for the coverage amount being below \$250k, which is negative at the community level but positive at the county level. 
This difference arises because coarser levels of spatial aggregation can mask the localized relationships between the predictors and claim intensity. 
For example, there may be small pockets of higher coverage properties associated with higher claim intensities, inside counties whose overall exposure composition is dominated by lower coverage properties. When claims are observed at the community level, the model retains more localized information about where claims occur and can associate those higher claim areas with higher coverage amounts, a relationship that is reflected by a negative coefficient since it corresponds to the indicator of lower coverage. 
In contrast, when claims are only observed at the coarser county level, the model cannot distinguish whether claims arise from a localized area within the county. The county-level claim frequency is instead associated with the integrated intensity over the entire county, which may be dominated by lower coverage properties. As a result, the estimated association also reflects the county-wide exposure composition, leading to a coefficient that can differ in sign from the finer community level. Thus, when claim counts are observed at coarser levels, localized relationships can be obscured, especially for predictors whose spatial distribution is heterogeneous within the grouped regions.

Table \ref{tab:AICBIC} presents the AIC and BIC for model selection based on the in-sample storms, where the proposed point process-derived method that incorporates more granular predictors is favored over the MZINB regression counterpart with spatially aggregated predictors, at both the county and community levels. Full parameter estimates for the MZINB regressions with aggregated predictors are provided in the \hyperref[sec:Appendix]{Appendix}. Furthermore, comparing the total deviance in Table \ref{tab:AICBIC} for the goodness-of-fit of each model, we observe that the proposed method has better fit on the in-sample storms compared to the multivariate count regression, for spatially grouped claim counts at both the county and community levels. Note that the county-level and community-level columns are not directly comparable to each other due to the different dimensions of the storm-level multivariate claim count outcomes.

\begin{table}[h!] 
	\centering
	\singlespacing
	\caption{AIC, BIC, and total deviance for the proposed zero-inflated point process-derived model and the grouped multivariate zero-inflated negative binomial regression for in-sample storms.}  
	\resizebox{\textwidth}{!}{
	\begin{tabular}{lrrrrr}
		\toprule
		& \multicolumn{2}{c}{\textbf{County-level}} &       & \multicolumn{2}{c}{\textbf{Community-level}} \\
		\cmidrule{2-3}\cmidrule{5-6}          & \multicolumn{1}{c}{\textbf{ZI point process}} & \multicolumn{1}{c}{\textbf{MZINB regression}} &       & \multicolumn{1}{c}{\textbf{ZI point process}} & \multicolumn{1}{c}{\textbf{MZINB regression}} \\
		\midrule
		AIC   & 11,767 & 20,215 &       & 25,214 & 29,543 \\
		BIC   & 11,841 & 20,288 &       & 25,287 & 29,616 \\
		Deviance & 7,508 & 15,966 &       & 18,695 & 22,949 \\
		\bottomrule
	\end{tabular}%
	}
	\label{tab:AICBIC}
\end{table}%

\begin{figure}[h!]
	\centering
	\includegraphics[width=0.65\linewidth]{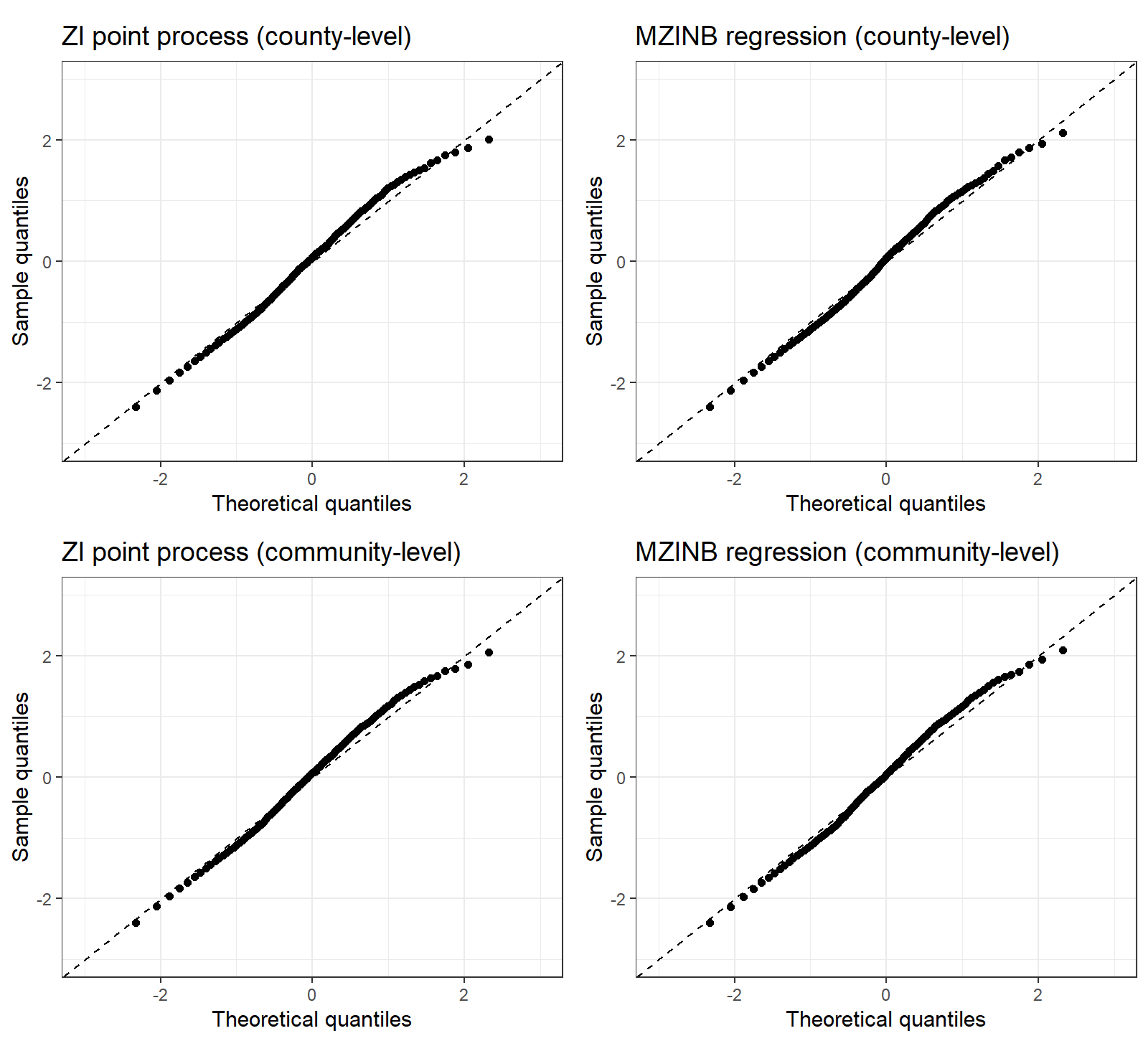}
	\caption{Q-Q plots of the Cox-Snell residuals of storm-level claim counts for the ZI point process (left) and MZINB regression (right), at the county (top) and community (bottom) levels of grouped outcomes.}
	\label{fig:nonrandPIT}
\end{figure}

To further assess the goodness-of-fit of the models on the in-sample storms, we examine the Cox-Snell residuals \citep{cox1968general} based on the probability integral transform (PIT) of the response using the fitted distributions. We consider the non-randomized version of the PIT for count data \citep{Czado2009}, applied to the storm-level claim frequency. Figure \ref{fig:nonrandPIT} displays the Q-Q plots of the Cox-Snell residuals for the zero-inflated point process model (left) and the multivariate zero-inflated negative binomial regression (right), fitted on spatially grouped losses at the county (top) and community (bottom) levels. The Q-Q plots follow the 45-degree line, suggesting that both models achieve a satisfactory distributional fit. 

\begin{figure}[H]
	\centering
	\includegraphics[width=0.65\linewidth]{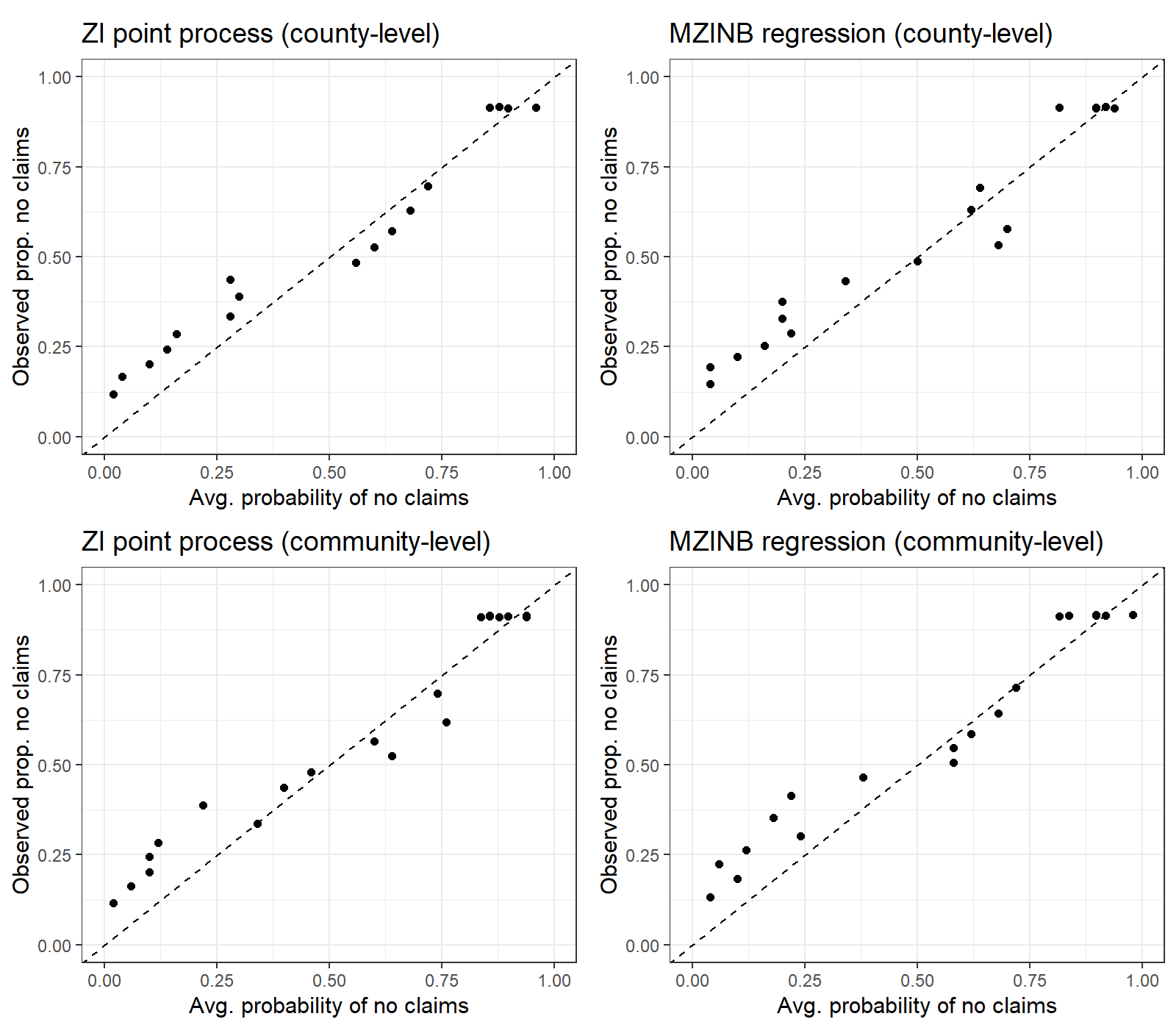}
	\caption{Reliability plots comparing the storm-level fitted probabilities of zero claims and the observed proportion of storms with zero claims, for the ZI point process (left) and MZINB regression (right), at the county (top) and community (bottom) levels of grouped outcomes. Storms are grouped into 20 bins by quantiles of fitted probabilities.}
	\label{fig:ZIreliab}
\end{figure}

We also assess the calibration of the zero-inflation in the storm-level claim frequency by examining the reliability curve for the probability of zero claims \citep{degroot1983comparison}. Figure \ref{fig:ZIreliab} plots the average fitted probabilities of zero claims for storms within 20 quantile-based bins along the $x$-axis, along with the corresponding conditional observed relative frequencies of storms with zero claims along the $y$-axis. The points are scattered around the 45-degree line for the proposed (left) and comparison (right) models at both county (top) and community (bottom) grouped outcome levels, indicating that the mechanism for modeling the observed and unobserved storm heterogeneity in the joint zeros effectively reflects the observed excess zeros in claim frequency.

\subsection{Prediction}

\begin{figure}[h!]
	\centering
	\includegraphics[width=0.85\linewidth]{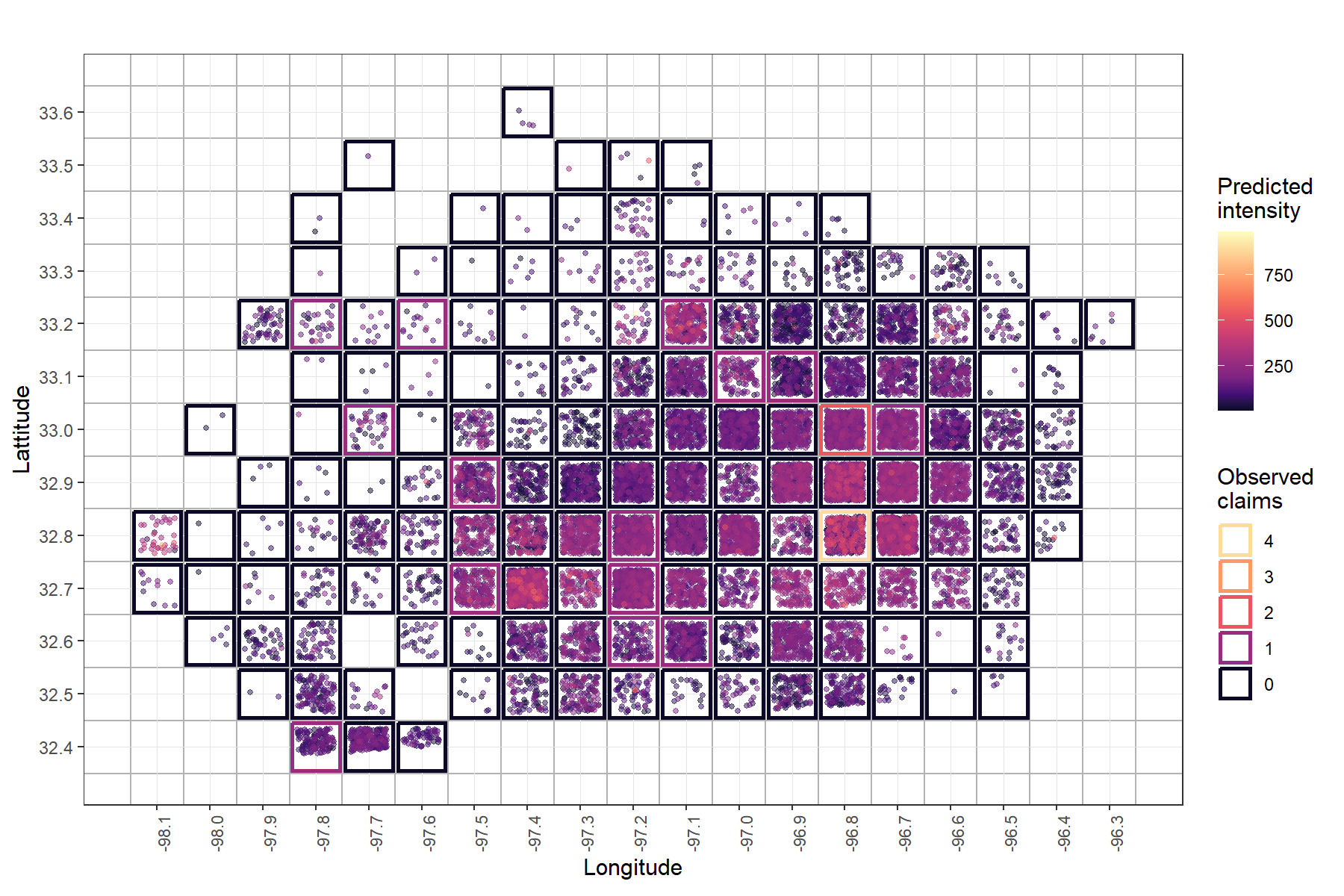} 
	\caption{Predicted zero-inflated spatial point process claim intensity at exposure locations for an out-of-sample storm, with observed community-level claim frequencies indicated by colored squares.}
	\label{fig:IntensityVsObserved}
\end{figure}

Even though the point process-derived model and the spatially grouped multivariate count regression can both be viewed as achieving adequate distributional fit for the storm-level claim frequency, the inclusion of the granular weather and exposure information provides additional localized predictive insights. Assuming the zero-inflated spatial point process as the underlying generating process for claims within a storm, we can predict the claim intensity for new storms, as well as obtain the joint predictive distributions of claim counts from any disjoint subregions within the storm window. As an example, based on the model estimated using the community-level grouped loss data, Figure \ref{fig:IntensityVsObserved} plots the predicted intensity of the zero-inflated point process for an out-of-sample storm at the property exposure locations. The actual observed community-level claim frequencies are indicated by the colored squares, where areas of relatively higher predicted intensities tend to correspond with areas with observed claims.

We more formally evaluate predictive performance in terms of both the point predictions and the probabilistic forecasts at the subregion level for the out-of-sample storms. 
We consider standard point prediction metrics such as the mean absolute error (MAE), root mean square error (RMSE), and Spearman correlation between the predicted and actual claim counts in a subregion of a storm. In addition, we evaluate the predictive distributions of the subregional claim counts, which are zero-inflated negative binomial under both the point process-derived model and the MZINB regression, through proper scoring rules. 
Scoring rules assign a numerical score to each observed outcome and its predictive distribution as negatively oriented penalties to minimize, and proper scoring rules address both calibration and sharpness when comparing competing forecasts \citep{gneiting2007probabilistic}. 
In practice, considering a variety of proper scoring rules allows us to take advantage of their differing emphases and strengths, and here we consider the quadratic or Brier score, spherical score, and ranked probability score adapted for count data \citep{Czado2009}.

Table \ref{tab:predMarginal} presents the point prediction metrics and proper scores for the subregion level claim counts, reported as averages over the out-of-sample storms. 
The ZI point process-derived model outperforms the MZINB regression in both point prediction (with lower MAE, RMSE, and higher Spearman correlation with the observed losses), as well as in probabilistic prediction (with lower average scores), at both the county and community levels of grouped losses. These results suggest that incorporating the granular exposure and weather information through the underlying point process enhances prediction for spatially grouped storm losses, compared to using aggregated predictors, even when the models induce the same  overdispersion and within-storm dependence structures.

\begin{table}[htbp]
	\centering
	\singlespacing
	\caption{Point prediction and probabilistic forecast performance for the county-level and community-level grouped losses in the out-of-sample storms for the ZI point process and MZINB regression.} 
	\resizebox{\textwidth}{!}{
	\begin{tabular}{lrrrrr}
		\toprule
		& \multicolumn{2}{c}{\textbf{County-level}} &       & \multicolumn{2}{c}{\textbf{Community-level}} \\
		\cmidrule{2-3}\cmidrule{5-6}          & \multicolumn{1}{c}{\textbf{ZI point process}} & \multicolumn{1}{c}{\textbf{MZINB reg.}} &       & \multicolumn{1}{c}{\textbf{ZI point process}} & \multicolumn{1}{c}{\textbf{MZINB reg.}} \\
		\midrule
		Point prediction &       &       &       &       &  \\
		\qquad MAE & 1.304 & 2.044 &       & 0.596 & 0.710 \\
		\qquad RMSE & 1.889 & 2.527 &       & 0.833 & 0.928 \\
		\qquad Spearman & 0.617 & 0.586 &       & 0.620 & 0.590 \\ \addlinespace
		Probabilistic forecast &       &       &       &       &  \\
		\qquad Quadratic score & -0.743 & -0.698 &       & -0.805 & -0.796 \\
		\qquad Spherical score & -0.848 & -0.839 &       & -0.897 & -0.895 \\
		\qquad Ranked prob. score & 0.679 & 0.887 &       & 0.317 & 0.334 \\
		\bottomrule
	\end{tabular}%
	}
	\label{tab:predMarginal}%
\end{table}%

In addition to evaluating the predictive distributions of the marginal subregion level claim counts, we also assess the overall joint predictive distributions from subregions of the same storm. Fewer multivariate scoring rules have been proposed in the literature compared to the univariate setting \citep{bjerregaard2021introduction}; we consider two complementary proper multivariate scoring rules. The energy score generalizes the ranked probability score to the multivariate setting to consider the overall joint predictive distribution, and is more sensitive to bias and variance discrepancy \citep{gneiting2007strictly}. Meanwhile, the variogram score targets the verification of the dependence structure, to which the energy score is less sensitive \citep{scheuerer2015variogram}. The energy and variogram scores are parameterized by orders $\alpha_{ES} \in (0, 2)$ and $\alpha_{VS}>0$ respectively, and we consider commonly used values $\alpha_{ES} \in \{0.5, 1, 1.5\}$ and $\alpha_{VS} \in \{ 0.5, 1, 2 \}$. The closed-form expressions for the scores are generally not tractable and we compute them from Monte Carlo approximation by drawing $1{,}000$ samples of $\bm{n}_i^* \in \mathbb{R}^{k_i}$ from the predictive distributions of each storm $i$. Table \ref{tab:predMultivariate} presents the proper multivariate scores for the ZI point process and MZINB regression forecasts, averaged over the out-of-sample storms, for joint count outcomes at both the county and community levels. Once again, the average scores are lower for the point process-derived model, indicating that incorporating more granular heterogeneity allows us to obtain more effective joint predictive distributions of the spatially grouped losses from a storm. 

\begin{table}[htbp]
	\centering
	\singlespacing
	\caption{Multivariate probabilistic forecast evaluation for the joint losses from a storm, grouped at the county and community levels, for the ZI point process and MZINB regression. Scores are reported as averages over the out-of-sample storms.}
	\begin{tabular}{lrrrrr}
		\toprule
		& \multicolumn{2}{c}{\textbf{County-level}} &       & \multicolumn{2}{c}{\textbf{Community-level}} \\
		\cmidrule{2-3}\cmidrule{5-6}          & \multicolumn{1}{c}{\textbf{ZI point process}} & \multicolumn{1}{c}{\textbf{MZINB reg.}} &       & \multicolumn{1}{c}{\textbf{ZI point process}} & \multicolumn{1}{c}{\textbf{MZINB reg.}} \\
		\midrule
		Energy score &       &       &       &       &  \\
		\qquad $\alpha_{ES}=0.5$ & 0.632 & 0.653 &       & 0.556 & 0.570 \\
		\qquad $\alpha_{ES}=1$ & 2.456 & 2.604 &       & 1.778 & 1.815 \\
		\qquad $\alpha_{ES}=1.5$ & 13.205 & 14.097 &       & 7.604 & 7.722 \\ \addlinespace
		Variogram score &       &       &       &       &  \\
		\qquad $\alpha_{VS}=0.5$ & 35    & 43    &       & 287   & 302 \\
		\qquad $\alpha_{VS}=1$ & 1,141 & 1,240 &       & 2,318 & 2,359 \\
		\qquad $\alpha_{VS}=2$ & 3,248,451 & 3,290,755 &       & 1,313,175 & 1,314,065 \\
		\bottomrule
	\end{tabular}%
	\label{tab:predMultivariate}%
\end{table}%

We confirm the superior predictive performance of the ZI point process over the MZINB regression using one-sided hypothesis tests. The Diebold-Mariano test of equal predictive performance between the two forecasts indicates that the ZI point process significantly outperforms the MZINB regression based on all scoring rules. In addition, a one-sided binomial test of equal proportions indicates that the ZI point process is preferred for more out-of-sample storms than the MZINB regression based on all scoring rules except the variogram score, where there is insufficient evidence to reject the null hypothesis. 
Table \ref{tab:predHypoTest} provides the Diebold-Mariano test statistics, as well as the observed proportion of out-of-sample storms where the ZI point process outperformed the MZINB regression, for both marginal and multivariate grouped loss outcomes. The corresponding $p$-values for the one-sided Diebold-Mariano and binomial tests are in parentheses. 
The proportions in Table \ref{tab:predHypoTest} further suggest that the point process model outperforms the MZINB regression more often when the MZINB predictors are aggregated at the coarser county level, compared to at the finer community level, for all scoring rules except the variogram score. This indicates that the more spatially aggregated the observed loss outcomes, the greater the value in incorporating the granular predictors in analyses, rather than aggregating them to the same level as the loss outcomes.  

\begin{table}[htbp]
	\centering
	\singlespacing
	\caption{Hypothesis tests comparing predictive performance of the ZI point process and the MZINB regression. Diebold-Mariano test statistics and the proportion of storms where the ZI point process outperformed the MZINB regression, with one-sided test $p$-values in parentheses.}
	\resizebox{\linewidth}{!}{
	\begin{tabular}{lrrrrrrrrr}
		\toprule
		\textbf{Scoring rule} & \multicolumn{4}{c}{\textbf{County-level}} &       & \multicolumn{4}{c}{\textbf{Community-level}} \\
		\cmidrule{2-5}\cmidrule{7-10}          & \multicolumn{2}{c}{\textbf{Diebold-Mariano}} & \multicolumn{2}{c}{\textbf{Binomial}} &       & \multicolumn{2}{c}{\textbf{Diebold-Mariano}} & \multicolumn{2}{c}{\textbf{Binomial}} \\
		\cmidrule{1-5}\cmidrule{7-10}    Quadratic & -9.57 & (<0.0001) & 76.7\% & (<0.0001) &       & -4.22 & (<0.0001) & 67.8\% & (<0.0001) \\
		Spherical & -8.62 & (<0.0001) & 65.2\% & (<0.0001) &       & -3.30 & (0.0005) & 55.2\% & (0.0323) \\
		Ranked probability & -4.14 & (<0.0001) & 77.9\% & (<0.0001) &       & -3.73 & (<0.0001) & 72.3\% & (<0.0001) \\
		Energy $(\alpha_{ES}=0.5)$ & -3.13 & (0.0009) & 72.0\% & (<0.0001) &       & -5.85 & (<0.0001) & 61.9\% & (<0.0001) \\
		Energy $(\alpha_{ES}=1)$ & -4.21 & (<0.0001) & 74.0\% & (<0.0001) &       & -5.85 & (<0.0001) & 63.1\% & (<0.0001) \\
		Energy $(\alpha_{ES}=1.5)$ & -3.94 & (<0.0001) & 77.0\% & (<0.0001) &       & -4.96 & (<0.0001) & 64.6\% & (<0.0001) \\
		Variogram $(\alpha_{VS}=0.5)$ & -3.16 & (0.0008) & 44.2\% & (0.9852) &       & -3.10 & (0.001) & 52.8\% & (0.1641) \\
		Variogram $(\alpha_{VS}=1)$ & -2.80 & (0.0025) & 43.1\% & (0.9955) &       & -2.84 & (0.0023) & 52.2\% & (0.2235) \\
		Variogram $(\alpha_{VS}=2)$ & -2.56 & (0.0052) & 42.2\% & (0.9984) &       & -2.48 & (0.0066) & 47.2\% & (0.8613) \\
		\bottomrule
	\end{tabular}%
	} 
	\label{tab:predHypoTest}%
\end{table}%

Our application of the zero-inflated mixed Poisson spatial point process on the real replicated grouped storm loss data reveals several valuable insights. 
First, incorporating densely measured property exposure and weather information to predict grouped loss outcomes more effectively addresses the localized heterogeneity of storms and storm losses, and achieves better fit compared to a multivariate count regression on aggregated predictors that similarly captures zero-inflation heterogeneity, within-storm dependence, and overdispersion characteristics. 
Second, even though a multivariate count regression 
can achieve a satisfactory distributional fit for claim frequency, making full use of the granular predictors available through an underlying point process provides more localized insights and improves the predictive performance, in both point and probabilistic predictions. Third, the more coarsely the grouped loss outcomes are observed, the more prediction-improving information is lost by aggregating the predictors to the same level. 

\section{Conclusion}\label{sec:Conclusion}

Severe convective storms constitute one of the most substantial yet challenging perils for property insurers in Canada and the United States. Predicting storm losses is complicated by pronounced heterogeneity, both at fine spatial scales within individual storms and across different storm events, as well as by the frequent occurrence of zero losses. 
Although increasingly rich sources of meteorological and property exposure data are available to improve loss prediction for these localized events, such information is typically aggregated in empirical analyses when loss responses are observed on a spatially grouped basis, for example, as claim counts at the county level. This aggregation results in the loss of potentially informative fine-scale detail. 

We introduce a latent zero-inflated mixed Poisson spatial point process as a framework for incorporating granular predictor information to analyze spatially grouped storm losses. Motivated by the underlying stochastic process of random claim locations arising from a storm, we derive the corresponding model for the observed multivariate zero-inflated count data that flexibly accommodates changing geographical segmentation across different storms. The model incorporates the effects of observed and unobserved storm heterogeneity in the joint zero claim counts, dependence between locations within the same storm, as well as the densely measured weather and property information to provide more localized prediction insights on the geographical distribution of claims from a storm. 

Through simulation studies, we demonstrate that the proposed zero-inflated spatial point process framework and its associated EM algorithm estimation procedure effectively recover the effects of unobserved storm heterogeneity and granular predictors on spatially grouped loss outcomes across varying levels of aggregation and perform comparably to the fully granular exact location setting. An application on real storm loss data further shows that incorporating granular weather and property information improves predictive performance for multivariate count outcomes, compared to using aggregated predictors, particularly when losses are observed more coarsely. 

While our study assumes a stationary underlying generating process of claims within a storm to focus on the integration of rich predictor information for spatially grouped losses, the potentially evolving storm, weather, and property exposure patterns suggest that future work may jointly consider the storm occurrence and the loss outcomes, including temporal trends and the compounding effect of multiple clustered storm events. 
As insurers continue to face the growing challenge of severe convective storms, leveraging rich weather and exposure datasets alongside traditional claims data enhances the characterization of storm heterogeneity to support more informed weather property risk management. 

\singlespacing

\section*{Acknowledgements}
The authors would like to thank the anonymous referees and the Editor for their thoughtful comments and suggestions that improved the paper. 
Lisa Gao acknowledges financial support from the Natural Sciences and Engineering Research Council of Canada (RGPIN-03389-2023). Tianxing Yan acknowledges support from the Society of Actuaries James C.\@ Hickman Scholars Program.

\section*{Competing interests}
The authors have no potential competing or conflicting interests to declare.

{\small
\bibliography{References}
\bibliographystyle{chicago}
}

\appendix 
\renewcommand{\thetable}{A\arabic{table}}
\renewcommand{\thealgocf}{A\arabic{algocf}}
\renewcommand{\thefigure}{A\arabic{figure}}
\setcounter{table}{0}
\setcounter{algocf}{0}
\setcounter{figure}{0}

\section*{Appendix}\label{sec:Appendix}

\begin{algorithm}[H]
	\singlespacing
	\SetAlgoLined
	\SetKwInOut{Input}{Input}
	\SetKwInOut{Output}{Output}
	\Input{window $W$, $\boldsymbol{\theta}=(\eta_0, \eta_1, \beta_1, \beta_2, \gamma, \psi)$}
	\Output{replicated grouped multivariate claim count outcomes $\bm{n}_i,\, i= 1, \ldots, m$}
	\BlankLine
	\For{$i=1, \ldots, m$}{
		Partition $W$ into subregions $B_1, \ldots, B_k$;\\
		Generate $u_i\sim Gamma$ with shape $1/\psi$ and scale $\psi$;\\
		Generate $z_i \sim N(0,1)$;\\
		Compute $p_i = (1+\exp\{ - (\eta_0 + z_i \eta_1) \})^{-1}$;\\
		
		Generate $v_i \sim U(0,1)$;\\
		\If{$v_i>p_i$}{
			Generate location $s_i$ uniformly within $W$;\\
			Compute $x_{i,1}(s) = (|s-s_i| + 0.1)^{-1}$;\\
			Generate $x_{i,2}(s) \sim U(0,1)$;\\
			Compute $\lambda_i^*=\max_{s \in W} \lambda_i(s|u_i)$, where $\lambda_i(s|u_i)=u_i \gamma \exp(x_{i,1}(s) \beta_1 
			+ x_{i,2}(s)\beta_2)$;\\
			Generate $n_i^* \sim Poisson(\lambda_i^*)$;\\
			\For{$j=1, \ldots, n_i^*$}{
				Generate location $y_{ij}^*$ uniformly within $W$;\\
				Generate $v_{ij}^* \sim U(0,1)$;\\
				Accept candidate $y_{ij}^*$ into point pattern $\bm{y}_i$ if $v_{ij}^* \le \lambda_i({y}_{ij}^*|u_i)/\lambda_i^*$;
			}
			\For{$l=1, \ldots, k$}{
			Count number of points from $\bm{y}_i$ in subregion $B_{l}$ as $n_{il}=\sum_{j=1}^{|\bm{y}_i|} 1\{ (\bm{y}_i)_j \in B_l \}$;\\
			}
			\Return{$\bm{n}_i = (n_{i1}, \ldots, n_{ik})$;}
		}
		\Else{$n_{i+}=0$}
	}
	\caption{Generation of multivariate grouped loss data from the underlying zero-inflated mixed-effects Poisson point process.}
	\label{alg:ZIPoissonpointprocess}
\end{algorithm}

\begin{table}[H]
	\centering
	\singlespacing
	\caption{Estimation results for the storm-level multivariate zero-inflated negative binomial regression, for losses and predictors grouped at the county level and the community level.}
	\resizebox{\textwidth}{!}{
		\begin{tabular}{l r r l r r l}
			\toprule
			\textbf{Parameter} 
			& \multicolumn{3}{c}{\textbf{County-level}} 
			& \multicolumn{3}{c}{\textbf{Community-level}} \\
			\cmidrule{2-4}\cmidrule{5-7}
			& \textbf{Est.} 
			& \textbf{Std.\@ error} 
			& 
			& \textbf{Est.} 
			& \textbf{Std.\@ error}
			&  \\
			\midrule
			
			Zero-inflation coefficients & & & & & & \\
			\qquad Intercept & 0.339 & 1.495 & & 0.552 & 1.680 & \\
			\qquad log number of exposures & -0.754 & 0.086 & *** & -0.793 & 0.098 & *** \\
			\qquad log area & -0.691 & 0.165 & *** & -0.568 & 0.183 & ** \\
			\qquad log max precipitation & 0.032 & 0.115 & & 0.063 & 0.129 & \\
			\qquad log max wind speed & 0.614 & 0.565 & & 0.586 & 0.643 & \\ 
			\addlinespace
			
			Multivariate negative binomial coefficients & & & & & & \\
			\qquad Intercept & 0.002 & 0.001 & ** & $<0.001$ & $<0.001$ & * \\
			\qquad Average log precipitation & 0.303 & 0.025 & *** & 0.353 & 0.023 & *** \\
			\qquad Average log wind speed & 2.116 & 0.125 & *** & 1.680 & 0.114 & *** \\
			\qquad Average log building age & 0.498 & 0.038 & *** & 1.004 & 0.033 & *** \\
			\qquad Flood zone risk & & & & & & \\
			\qquad \qquad Proportion in High & 0.108 & 0.063 & & 0.525 & 0.059 & *** \\
			\qquad \qquad Proportion in Moderate & 4.028 & 0.290 & *** & 1.654 & 0.167 & *** \\
			\qquad \qquad Proportion in Low & -0.925 & 0.092 & *** & -0.014 & 0.083 & \\
			\qquad Proportion with coverage under \$250k & -0.298 & 0.063 & *** & -0.493 & 0.060 & *** \\
			\qquad Proportion elevated & -1.077 & 0.128 & *** & -0.657 & 0.090 & *** \\
			\qquad $\psi$ & 2.435 & 0.182 & *** & 2.805 & 0.196 & *** \\
			\bottomrule
			\multicolumn{7}{l}{\footnotesize *$p<0.05$, **$p<0.01$, ***$p <0.001$.} \\
		\end{tabular}%
	}
	\label{tab:ZINBregResults}%
\end{table}%

\end{document}